\documentclass{aastex63}

\usepackage{amsmath, bm}

\received{xx xx, xxxx}
\revised{xx xx, xxxx}
\accepted{xx xx, xxxx}
\submitjournal{AJ}

\shorttitle{Planetesimal Dynamics in the Presence of a Giant Planet}
\shortauthors{Guo and Kokubo}

\begin{document}
\title{Planetesimal Dynamics in the Presence of a Giant Planet}

\correspondingauthor{Kangrou Guo}
\email{carol.kwok@grad.nao.ac.jp}

\author{Kangrou Guo}
\affiliation{The University of Tokyo}
\affiliation{National Astronomical Observatory of Japan}

\author{Eiichiro Kokubo}
\affiliation{The University of Tokyo}
\affiliation{National Astronomical Observatory of Japan}

\begin{abstract}
Standard models of planet formation explain how planets form in axisymmetric, unperturbed disks in single star systems.
However, it is possible that giant planets could have already formed when other planetary embryos start to grow.
We investigate the dynamics of planetesimals under the perturbation of a giant planet in a gaseous disk.
Our aim is to understand the effect of the planet's perturbation on the formation of giant planet cores outside the orbit of the planet.
We calculate the orbital evolution of planetesimals ranging from $10^{13}$ to $10^{20}$g, with a Jupiter-mass planet located at 5.2 au.
We find orbital alignment of planetesimals distributed in $\simeq 9$-15 au, except for the mean motion resonance (MMR) locations.
The degree of alignment increases with increasing distance from the planet and decreasing planetesimal mass. 
Aligned orbits lead to low encounter velocity and thus faster growth.
The typical velocity dispersion for identical-mass planetesimals is $\sim \mathcal{O}(10)$ $\rm{m\text{ } s}^{-1}$ except for the MMR locations.
The relative velocity decreases with increasing distance from the planet and decreasing mass ratio of planetesimals.
When the eccentricity vectors of planetesimals reach equilibrium under the gas drag and secular perturbation, the relative velocity becomes lower when the masses of two planetesimals are both on  the larger end of the mass spectrum.
Our results show that with a giant planet embedded in the disk, the growth of another planetary core outside the planet orbit might be accelerated in certain locations.

\end{abstract}

\section{introduction} \label{sec:intro}

In the classic \textit{core accretion} model of giant planets (e.g., \citet{Mizuno_1980}, \citet{Pollack_et_al_1996}), a massive planetary core first forms, followed by inflow of nebular gas onto the growing core as the accreted envelope cools and contracts. 
When the envelope mass becomes large enough,  its gravity overcomes its pressure gradient, triggering rapid gas accretion (runaway gas accretion) that forms the gas giant planet before the gaseous disk disperses. 
However the core accretion model often suffers from a timescale problem, in that the core might not be able to reach the critical mass early enough to start the runaway gas accretion before the nebular gas disperses.

Perturbations from a massive body in the disk might aid in accelerating the formation of a planetary core.
Assuming that Jupiter and Saturn formed towards the end of the Sun's formation and possibly prior to the quiescent stage of disk evolution, \citet{KW00} investigated the early evolution of planetesimal orbits in the inner Solar System under the perturbation of these two gas giants. 
They found that the coupling effect of the gas drag and the gravitational perturbation from Jupiter and Saturn introduces a size-dependent phasing of the planetesimal orbital elements or orbital alignment, which leads to a pronounced dip in the encounter velocities between planetesimals of similar sizes. 
For low encounter velocities, a colliding pair of planetesimals are more likely to merge than disrupt, and thus the growth rate of the planetary core is enhanced \citep{Kortenkamp_et_al_2001}. 
Also, for lower encounter velocities among swarms of planetesimals, planetary bodies can form in environments where planetesimal orbits have higher eccentricities and inclinations than are usually considered \citep{Kortenkamp_et_al_2001}. 
The effect of secular perturbation on planetesimals has also been studied for binary star systems.
\citet{Marzari_Scholl_2000} investigated the evolution of planetesimals orbiting the primary star of Alpha Centauri and found that the secular perturbation from the secondary star forces an efficient pericenter alignment between neighboring orbits of planetesimals; meanwhile, gas drag dampens orbital eccentricities and restores a strong pericenter alignment for equal-size bodies.
\citet{THEBAULT2006193} further explored the relative velocity among planetesimals in circumprimary disks and its dependence on the binary parameters.

The studies mentioned above all focused on planetesimals orbiting between the central star and the perturber.
\citet{KW00} and \citet{Kortenkamp_et_al_2001} modeled the inner disk interior to Jupiter's orbits. \citet{Marzari_Scholl_2000} and \citet{THEBAULT2006193} focused on planetesimals in a circumprimary disk.  

However, it is equally important to explore the dynamics of planetesimals orbiting exterior to the orbit of the perturber itself. 
For example, one application is the formation of P-type planets around binary stars. 
The conditions for planetesimal accretion in a circumbinary disk have been investigated by \citet{Moriwaki_Nakagawa_2004}. 
They identified the region where planetesimal accretion can occur under the secular perturbation of the stellar companion. 
However, they neglected the effect of gas drag and the mutual perturbation of planetesimals (self-gravity). 
The effect of secondary perturbation and gas drag on the relative velocities of planetesimals has been investigated in e.g.,  \citet{Scholl_et_al_2007}  and \citet{Marzari_et_al_2008}. 
\citet{Scholl_et_al_2007} focused on the distribution of encounter velocities between planetesimals in the 0.5-100 km-size range. 
For each binary configuration (binary mass ratio and eccentricity), they derived the critical radial distance from the binary barycenter beyond which accretion is possible.
\cite{Marzari_et_al_2008} used a rather complicated model of the gas component of the disk: they included a hybrid model where the gas drag and the gas disk gravity for solid component were derived from hydrodynamical simulations. 
While confirming that pericenter alignment of planetesimals is a robust result, they found that the pericenters of planetesimals are less collimated compared to early predictions from stationary and axisymmetric disk approximation.
The self-gravity of planetesimals was again neglected in both of these studies.
Apart from forming planets in circumbinary disks, another applicable case is the formation of another giant planet core exterior to the perturber, assuming that the perturber is a giant planet. 
For example, since Jupiter is much more massive than Saturn, it is reasonable to assume that Jupiter completed formation before Saturn. 
If Saturn's core started to form while Jupiter already existed, it would have been subject to Jupiter's perturbation through-out the formation process. 
So far, there has been few studies investigating the impact of a giant planet on planetesimal accretion in disk regions \textit{exterior} to its orbit.

In this work, we consider the outer disk (exterior to the perturber's orbit) and aim at finding out the impact of the perturbation from a giant planet on the dynamical evolution of planetesimals in this region.
In particular, we examine whether Jupiter's perturbation,  together with nebula gas drag, induces any accretion-friendly features on the orbits of planetesimals in the outer disk that would have lowered their random encounter velocities and thereby accelerated the formation of Saturn's core. 
Such a scenario might also be applicable for extrasolar systems with two giant planets in similar configuration as Jupiter and Saturn.
In future work, we will consider a more realistic scenario by introducing the self-gravity of planetesimals to this model. Building on that, we plan to extend the framework of this study to binary systems, i.e., how the perturbation from a stellar companion affects the dynamics of planetesimals in a circumbinary disk.

The paper is arranged as follows: in Section \ref{sec:intro}, we briefly summarize the previous studies related to the topic of this work. In Section \ref{sec:method}, we explain our model setup and the calculations of orbits and relative velocity. In Section \ref{sec:basic_dynamics}, we analytically show how the orbit of a planetesimal evolves using the secular perturbation theory. Section \ref{sec:results} presents a detailed illustration of our results. Finally, we summarize our paper and discuss our future perspectives in Section \ref{sec:summary}.

\section{Method} \label{sec:method}

\subsection{Model setup}

We consider a system with a central star of mass $M_* = 1 M_{\odot}$ and a protoplanetary disk comprised of gas and planetesimals. 
For the gas component, we employ the classical Hayashi model \citep{Hayashi_1981}:
\begin{equation}
\Sigma_{\rm{g}} = 2.4 \times 10^3 \left(\frac{r}{\rm{au}}\right)^{-3/2} \text{ g cm}^{-2}.
\end{equation}
This corresponds to a 50\% more massive disk than the minimum mass solar nebula (MMSN).
The spatial density at the midplane is
\begin{equation}
    \rho_{\rm{gas}} = 2.0 \times 10^{-9} \left( \frac{a}{1 \rm{AU}} \right)^{-11/4} \rm{g}\text{ }\rm{cm}^{-3}.
\end{equation}

We put a planet with mass $M_{\rm{p}} = 1 M_{\rm{Jup}}$ at 5.2 au with an orbital eccentricity of 0.048, which is the same as Jupiter. 
For simplicity, we set the planet's inclination to be 0 and its orbital pericenter on the $x$-axis ($\omega = 0$).

As for the solid surface density, we distribute the planetesimals in an annulus of 8-15 au, following a power law distribution proportional to $a^{-3/2}$:
\begin{equation}
\Sigma_{\rm{d}} = 42 \left(\frac{a}{1\text{ au}}\right)^{-3/2} \text{ g cm}^{-2}.
\label{eq:Sigma_d}
\end{equation}
However, for each planetesimal mass, we fix the number of particles $N$ so that the actual distribution of solid surface density follows Eq. (\ref{eq:Sigma_d}) with a varying factor.
The initial orbital eccentricities and inclinations of the planetesimals follow the Rayleigh distribution with dispersion $\sigma_e = \sigma_i = 2r_{\rm{H}}/a$, where $r_{\rm{H}}$ is the Hill Radius of the planetesimal given by
\begin{equation}
r_{\rm{H}} = \left(\frac{2m}{3M_*}\right)^{1/3} a.
\end{equation}
The Hill radius is the radius of the gravitational potential well of a body in the rotating frame \citep[e.g.,][]{kokubo2012dynamics}. 
The other orbital elements ($\Omega$, $\omega$, $\tau$) are randomly chosen from 0 to $2 \pi$. 
Since the planetesimals are in the outer disk exterior to the snow line, we set their density to be 1 $\text{g} \, \text{cm}^{-3}$. 
The gas drag force is calculated by the formula in \citet{Adachi_et_al_1976}:
\begin{equation}
f_{\rm{gas}} = \frac{1}{2}C_{\rm{D}} \pi r^2 \rho_{\rm{gas}} u^2,
\end{equation}
where $C_{\rm{D}}$ is the non-dimensional drag coefficient, $r$ is the particle radius, and $u$ is the relative velocity between the gas and the particle. 
Since the mass range of planetesimals we chose corresponds to a radius range of roughly 0.1-30 km, we adopt the drag law with a quadratic dependence on the relative velocity $u$ and a value of $C_{\rm{D}} = 1$ (see Fig. 4 in \citet{Adachi_et_al_1976} for the value range of $C_{\rm{D}}$).
The gas drag timescale for eccentricity is calculated by \citep{kokubo2012dynamics}
\begin{equation}
\tau_{\rm{drag},e} = \frac{e}{|de/dt|} \sim 10^5 \left(\frac{e}{0.1}\right)^{-1} \left(\frac{m}{M_{\oplus}}\right)^{1/3} \left(\frac{\rho}{3\rm{g cm}^{-3}}\right)^{2/3} \left(\frac{a}{1 \rm{AU}}\right)^{13/4} \rm{year},
\label{eq:tau_gas}
\end{equation}
where the effect of the sub-Keplerian rotation of gas $\eta$ is neglected for simplicity. 

The equation of motion of a planetesimal $i$ is given by
\begin{equation}
\frac{{\rm{d}}  \bm{v}_{\rm{i}}}{{\rm{d}}  t} = -GM_* \frac{\bm{x}_i}{|\bm{x}_i|^3} + GM_{\rm{p}} \frac{\bm{x}_{\rm{p}} - \bm{x}_i}{|\bm{x}_{\rm{p}} - \bm{x}_i|^3} + \bm{f}_{\rm{gas}}, 
\end{equation}
where the subscript "p" denotes quantities for the planet, which resembles Jupiter. 
The second term on the right-hand side describes the planet's perturbation on the particle. 
Following previous studies (e.g. \citet{Kortenkamp_et_al_2001}, \citet{Moriwaki_Nakagawa_2004}, and \citet{Scholl_et_al_2007}), we currently neglect the self-gravity of the planetesimals for simplicity.
We also do not consider the growth or fragmentation of planetesimals. In other words, the mass of the planetesimals remains the same throughout the simulation.

The initial conditions including the orbital elements of the planet and the planetesimals (definitions given in the table caption) are summarized in Table \ref{tab:model}.

\begin{deluxetable*}{C | C C}
\tablenum{1}
\tablecaption{Initial model setup\label{tab:model}}
\tablewidth{0pt}
\tablehead{\colhead{ } & \colhead{Planet} & \colhead{Planetesimal}}
\startdata
 N & 1 & 29534 \times 8 \\
m  & 0.001 \, M_{\odot} (1 \, M_{\rm{Jup}}) & 10^{13} - 10^{20} \, \rm{g} \\
a\, \text{(au)} & 5.2 & 8 - 15 \\
e & 0.048 & \text{Rayl.} (\sigma_e = 2r_{\rm{H}}/a) \\
i \, \text{(rad)} & 0 & \text{Rayl.} (\sigma_i = 2r_{\rm{H}}/a) \\
\Omega \, \text{(rad)} & 0 & [0, 2\pi] \\
\omega \, \text{(rad)} & 0 & [0, 2\pi] \\ 
\tau \, \text{(rad)} & 0 & [0, 2\pi] \\
\rho \, (\rm{g}\cdot \rm{cm}^{-3}) & - & 1 \\
\text{integration time (yr)} & \multicolumn{2}{C}{\simeq 3\times 10^6} \\
 \enddata
\tablecomments{This table shows the model setup and initial conditions. The rows are particle number, mass, semi-major axis, eccentricity, inclination, longitude of ascending node, argument of pericenter, time of pericenter passage, density, and time of integration. "Rayl." means that the initial distribution follows the Rayleigh distribution with dispersion $\sigma_e$ or $\sigma_i$.}
\end{deluxetable*}

\subsection{Numerical integrator}

We numerically investigate the spatial distribution and orbital evolution of planetesimals under the effect of nebula gas drag and perturbation from a giant planet. 
For the simulations, we use a fourth-order Hermite integrator \citep{Kokubo_Makino_2004} with block timesteps \citep{Makino_1991PASJ}. 
The Hermite scheme is a predict-evaluate-correct (PEC) scheme in which the acceleration and its time derivative are used to construct the interpolation polynomials of the acceleration \citep{Kokubo_et_al_1998}. 
The integrator optimized for planetary \textit{N}-body problems because it allows individual timesteps for close encounters to achieve high accuracy, while preserving the approximate time-symmetry \citep{Kokubo_et_al_1998, Kokubo_Makino_2004}. 
The individual timesteps are controlled by a hierarchical timestep algorithm where timesteps are "quantized" to powers of two (block timestep) \citep{Makino_1991PASJ}. 
More details about the integrator can be found in Sections 2 and 3 in \citet{Kokubo_Makino_2004}.

We integrate the system for $\simeq 3 \times 10^6$ years, a timescale long enough to account for the effect of secular perturbation and gas drag for all particle masses (see Section \ref{sec:timescales}  for more detailed discussion on timescales).

\subsection{Pericenter alignment}

\begin{figure*}
\gridline{\fig{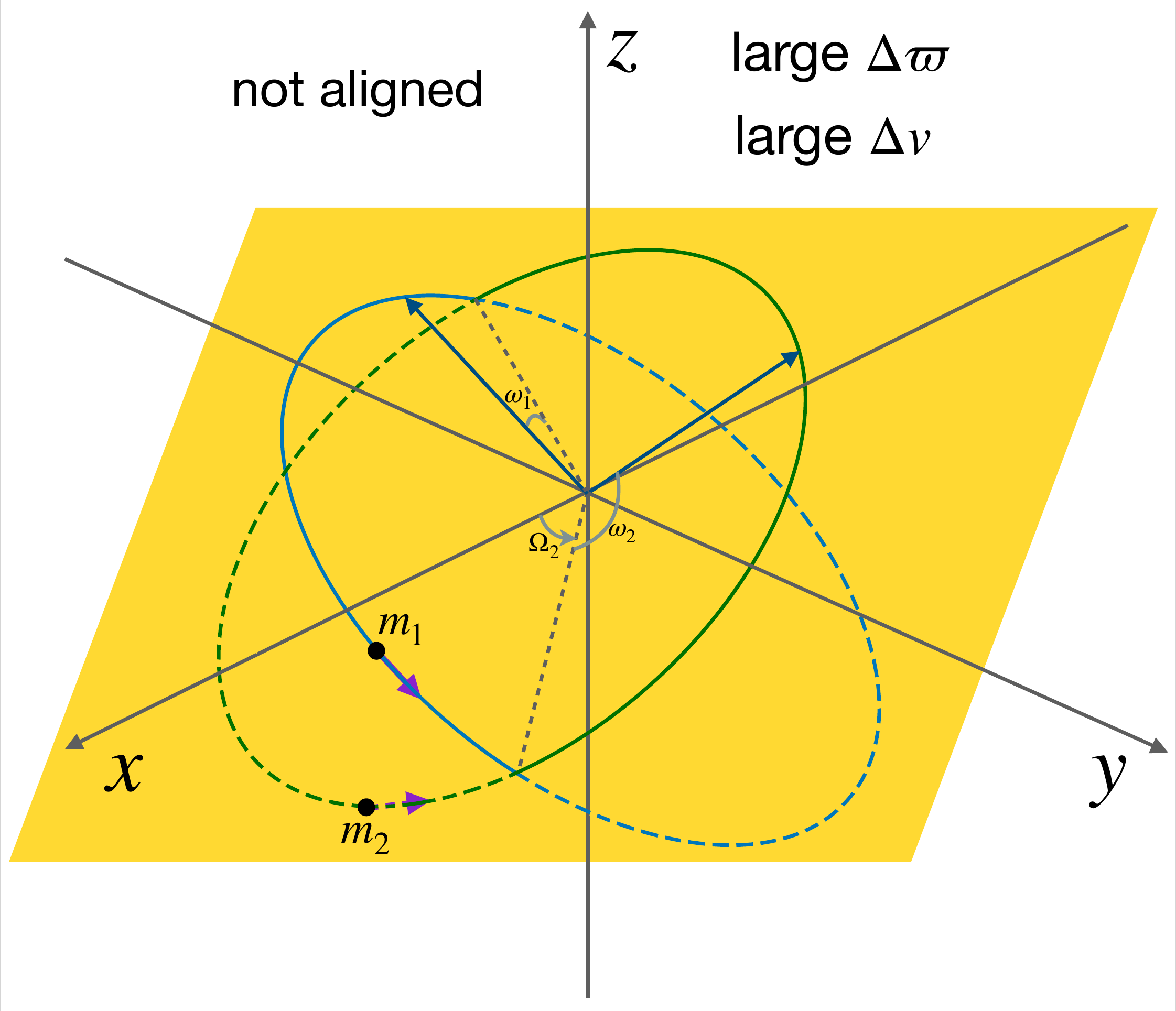}{0.5\textwidth}{(a)}
	\fig{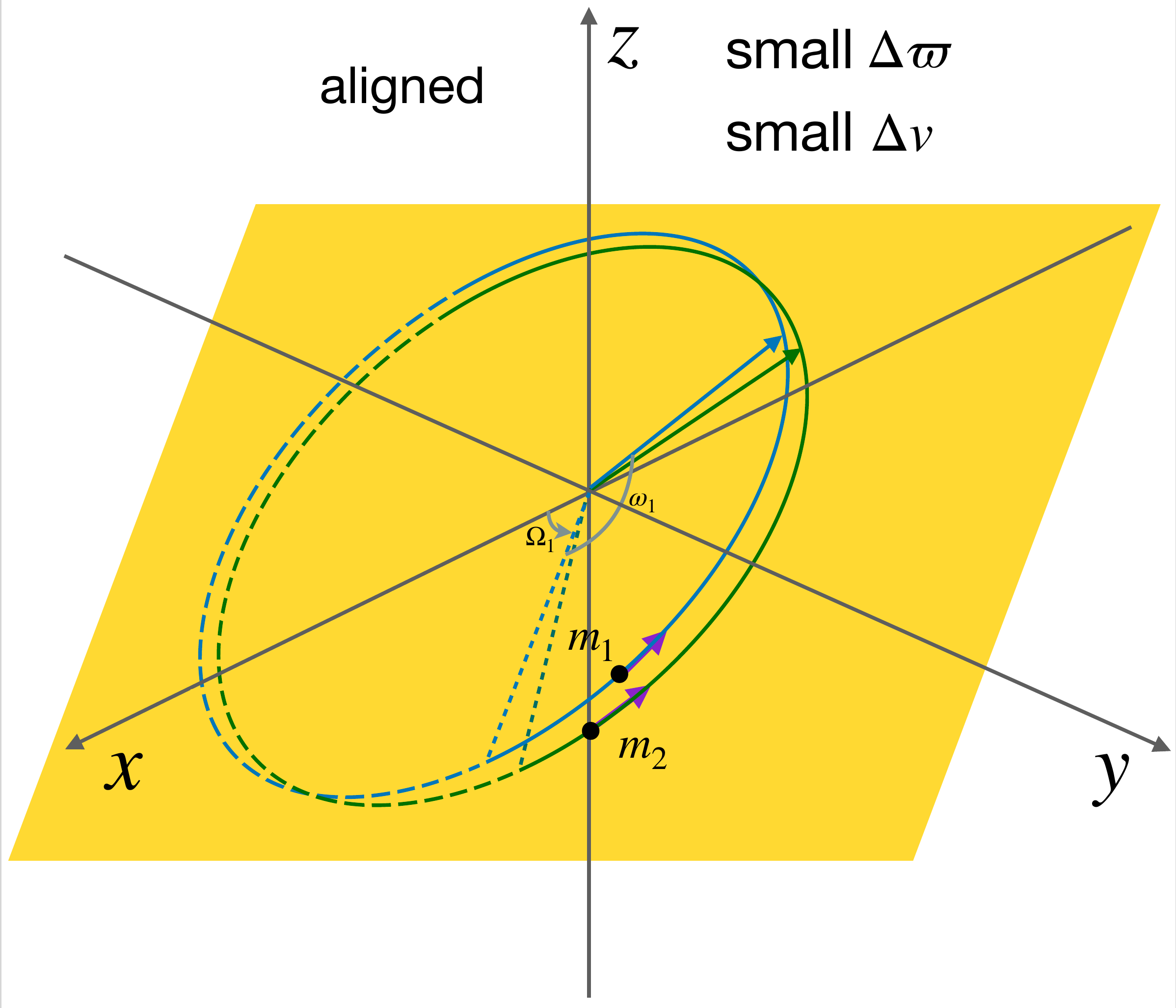}{0.5 \textwidth}{(b)}}
\caption{Schematic illustration of pericenter alignment. The dashed curves indicate those parts of the orbits that are below the reference plane (the yellow shaded plane). Plot (a): non-alignment case; $m_1$ and $m_2$ encounter with high relative velocity. Plot (b): alignment case; $m_1$ and $m_2$ encounter on almost tangential orbits and therefore with low relative velocity even when they are on eccentric and inclined orbits.}
\label{fig:apse_alignment}
\end{figure*}

The configuration of the planetesimal orbits has a direct impact on the random encounter velocities. 
One important characteristic of an orbit is its eccentricity vector.
The eccentricity vector is also known as the \textit{Laplace-Runge-Lenz vector}. Its direction points from the central star to the pericenter, and its magnitude is equal to the orbit's scalar eccentricity. 
The vector can be expressed with orbital elements by 
\begin{equation}
\bm{e} = e  \left( 
\begin{array}{c}
\cos{\Omega} \cos{\omega} - \sin{\Omega} \sin{\omega} \cos{i}\\
\sin{\Omega} \cos{\omega} + \cos{\Omega} \sin{\omega} \cos{i}\\ 
\sin{\omega} \sin{i}
\end{array} \right).
\end{equation}
When the inclination $i$ is very small, we can reduce the vector to two dimensions ($e_x$, $e_y$) as 
\begin{equation}
\bm{e} = e(\cos{\varpi}, \text{ } \sin{\varpi}),
\label{eq_eccvec}
\end{equation}

where $\varpi = \omega + \Omega$ is the longitude of pericenter. 
This is an indicator of the degree of pericenter alignment.
If two orbits are aligned, the difference in their longitudes of pericenter $\Delta \varpi$ is small, and the particles encounter each other on almost tangential trajectories with low relative velocities. 
Fig. \ref{fig:apse_alignment} is a schematic illustration of how the pericenter alignment of orbits affects the mutual encounter velocity of two particles.

\subsection{Encounter velocities}

The random encounter velocity among planetesimals is the most important factor that determines the growth rate of planetary bodies. 
It is a critical parameter that measures the potentiality of a planetesimal swarm to accumulate into larger bodies, as it determines whether mutual impacts result in accretion or disruption.
Generally, lower random encounter velocity indicates higher accretion rate.
Tracking the actual mutual encounter velocities (relative velocities) between colliding particle pairs during the simulation, however, is very time-consuming. 
Therefore, we statistically calculate the velocity dispersion for a population of identical-mass particles, and the relative velocity between two populations of different masses. 

For identical-mass particles, we take bins by semi-major axis and calculate the velocity dispersion for particles in these bins. 
The bin width is set to be $\sim 0.1$ au. 
We calculate the mean of the eccentricity vectors $(\bar{e}_x, \bar{e}_y)$ (i.e., the mean location of the pericenters) and the Keplerian velocity $v_{\rm{K}}(a)$ of a bin. 
The pericenter locations $(e_x, e_y)$ are given by Equation (\ref{eq_eccvec}), and the mean Keplerian velocity is 
\begin{equation}
v_{\rm{K}} = \bar{a} \sqrt{\frac{GM_*}{\bar{a}^3}}.
\label{eq_vk}
\end{equation}
Here $\bar{a}$ is the mean value of the semi-major axis of the particles in a bin.
Then the velocity dispersion $\sigma_v$ of $N$ particles in a bin is given by
\begin{equation}
\sigma_v \simeq \sigma_e v_{\rm{K}},
\label{eq:v_disp}
\end{equation}
where the dispersion of eccentricity $\sigma_e$ is given by
\begin{equation}
\sigma_e = \sqrt{\sigma_{e_x}^2+\sigma_{e_y}^2} = \sqrt{\frac{\sum_{i=1}^N (e_{x, i} - \bar{e}_x)^2}{N} + \frac{\sum_{i=1}^N (e_{y, i} - \bar{e}_y)^2}{N}}.
\label{eq:sigma_e}
\end{equation}
For particles of different masses, we still exploit the relation in Eq. (\ref{eq:v_disp}) to calculate the encounter velocity, but instead of the dispersion of eccentricity, we calculate the difference between the mean locations of the eccentricity vectors of different mass groups on the $e_x$-$e_y$ plane, i.e. $v_{{\rm{rel}},{ij}} \simeq \Delta e_{ij} v_{\rm{K}}$. 
However, sometimes the dispersion of the eccentricity vectors of a single mass group can be larger than the difference between the mean locations of the eccentricity vectors of two mass groups, i.e. $\sigma_{e,i} > \Delta e_{ij}$ (see Fig. \ref{fig:ecc_vec_mid}).
In this case, we use the larger value comparing $\sigma_{e,i}$ and $\Delta e_{ij}$ to give a conservative estimate of the encounter velocity.
Specifically, we first group the particles in a given range of semi-major axis by mass, and then calculate their mean pericenter location $(\bar{e}_x, \bar{e}_y)$. 
For example, in a specific range of $a$, the relative velocity between particle groups $i$ and $j$ is
\begin{equation}
v_{rel,ij} \simeq v_{\rm{K}} \cdot \max{\left(\Delta e_{ij}, \frac{\sigma_{e,i} + \sigma_{e,j}}{2}\right)},
\end{equation}
where $\Delta e_{ij}$ is the difference between the two mean pericenter locations of particles of group $i$ and $j$ 
\begin{equation}
\Delta e_{ij} = \sqrt{ (\bar{e}_{x,i} - \bar{e}_{x,j})^2 +  (\bar{e}_{y,i} - \bar{e}_{y,j})^2 }.
\end{equation}

We also calculate the two-body surface escape velocity $v_{esc}$ between $m_1$ and $m_2$, and compare it with the relative velocity. The escape velocity is given by
\begin{equation}
v^2_{esc} = \frac{2G(m_1+m_2)}{R_1+R_2},
\label{eq:v_esc}
\end{equation} 
where $R_1$ and $R_2$ are the particle radii, $R = (3m/4\pi \rho)^{1/3}$.

After quantifying the pericenter alignment and random encounter velocities, we determine how they are influenced by the coupled effect of secular perturbation and gas drag.

\section{Basic Dynamics} \label{sec:basic_dynamics}

In this section, we analytically show how the pericenters of planetesimals become aligned under the coupled effect of secular perturbation and gas drag by using secular perturbation theory (e.g., \citet{Greenberg_1978, Marzari_Scholl_2000, Moriwaki_Nakagawa_2004}).

\subsection{Secular perturbation theory and the eccentricity vector}

The location of a pericenter is denoted as $(k, h)$ in the $e_x$-$e_y$ plane (i.e., the eccentricity vector), where $k$ and $h$ are defined as
\begin{eqnarray}
k &=& e\cos{\varpi}; \\
h &=& e \sin{\varpi}.
\end{eqnarray}

For the case where the perturber is within the orbit of the particle in motion ($r > r_{\rm{p}}$, as in our simulation case) and when the inclination is small, we can write the secular part of the time averaged disturbing function using the Legendre expansion as
\begin{equation}
\big \langle R\big \rangle = \mu  n_{\rm{p}}^2 \frac{a_{\rm{p}}^5}{a^3} \left[\frac{1}{4} \left( 1+\frac{3}{2}e^2\right) \left(1+\frac{3}{2}e_{\rm{p}}^2\right) + \frac{a_{\rm{p}}}{a}\frac{3}{8} e \left(-\frac{5}{2}e_{\rm{p}} - \frac{15}{8}e_{\rm{p}}^3\right) \cos{(\varpi - \varpi_{\rm{p}})} \right],
\end{equation}
where $n$ is the mean motion and $\mu$ is the mass ratio ($\mu = m_{\rm{p}} / (m_{\rm{p}} + M_*)$); again the subscript "p" denotes quantities of the planet, which resembles Jupiter in our simulations. We neglect the terms with $\sin(i)^2$ in the disturbing function. 
Here one concern is that the Legendre expansion of the disturbing function $\big \langle R \big \rangle$ assumes small values for the the semi-major axis ratio $a_{\rm{p}}/a$. 
For the case of the planet disturber here, in which $a_{\rm{p}}/a$ can be as large as $\simeq 0.5$ for $a=10$ au, this approximation of the disturbing function might not be very accurate. 
However, as we can see from comparing the locations of the equilibrium eccentricity vectors of the smallest masses in Fig. \ref{fig:hk_eq} (b) and Fig. \ref{fig:ecc_vec_mid} (e), the analytical results calculated from this approximated disturbing function do not significantly deviate from the data from numerical simulations. 
Therefore, we accept that the approximated disturbing function using Legendre expansion gives reasonable analytical predictions for our simulation data.

Plugging the disturbing function into the Lagrangian planetary equations, we can write the secular variations of $h$ and $k$ as 
\begin{eqnarray}
\frac{\text{d}h}{\text{d}t} &=& Ak-B \cos{\varpi_{\rm{p}}}, \label{eq:without_gas1}\\ 
\frac{\text{d}k}{\text{d}t} &=& -Ah + B \sin{\varpi_{\rm{p}}}, \label{eq:without_gas2}
\end{eqnarray}

where
\begin{eqnarray}
A &=& \mu \frac{3}{4} \frac{n^2_{\rm{p}}}{n} \left(\frac{a_{\rm{p}}}{a}\right)^5  \left( 1+\frac{3}{2} e^2_{\rm{p}}\right), \label{eq:A}\\
B &=& \mu \frac{3}{8} \frac{n^2_{\rm{p}}}{n} \left(\frac{a_{\rm{p}}}{a}\right)^6  \left(\frac{5}{2}e_{\rm{p}} + \frac{15}{8}e_{\rm{p}}^3\right).
\end{eqnarray}
The forced eccentricity imposed by the perturber can be calculated by
\begin{equation}
e_{\rm{forced}} = \frac{B}{A} = \frac{1}{2} \left(\frac{a_{\rm{p}}}{a} \right) \Big[\frac{(5/2)e_{\rm{p}} + (15/8)e_{\rm{p}}^3}{1+(3/2)e_{\rm{p}}^2} \Big].
\label{eq:forced_eccentricity}
\end{equation}

The solutions to Eq. (\ref{eq:without_gas1}) and Eq. (\ref{eq:without_gas2}) can be written as
\begin{eqnarray}
k &=& e_{\rm{prop}} \cos{(At+\varpi_{\rm{prop}})} + e_{\rm{forced}} \cos{\varpi_{\rm{p}}},\\
h &=& e_{\rm{prop}} \sin{(At+\varpi_{\rm{prop}})} + e_{\rm{forced}} \sin{\varpi_{\rm{p}}}.
\end{eqnarray}

From the solutions we can describe the evolution of the eccentricity vector in this way: if the particle is initially in a circular orbit and feels no gas drag, the proper eccentricity $e_{\rm{prop}} = |\bm{e} - \bm{e}_{\rm{forced}}|$ equals to $e_{\rm{forced}}$; its pericenter $(k,h)$ circulates the location of the forced eccentricity $(e_{\rm{forced}} \cos{\varpi_{\rm{p}}}, e_{\rm{forced}} \sin{\varpi_{\rm{p}}})$ with a radius $e_{\rm{forced}}$, and with an angular velocity $\text{d}\varpi_{\rm{prop}}/\text{d}t = A$ in the $e_x$-$e_y$ plane \citep{Moriwaki_Nakagawa_2004}.

The timescale for secular perturbation $\tau_{\rm{sec}}$ is defined as the circulation period of the pericenter locus $(e_x, e_y)$ around the forced eccentricity $(e_{\rm{forced}} \cos{\varpi_{\rm{p}}}, e_{\rm{forced}} \sin{\varpi_{\rm{p}}})$ on the eccentricity vector plane:
\begin{equation}
\tau_{\rm{sec}} = \frac{2\pi}{A}.
\label{eq:tau_sec}
\end{equation}

\subsection{Equilibrium of the eccentricity vector}

If the motion of the planetesimal is subjected to nebula gas drag, which dampens its eccentricity at a rate proportional to $e^2$ \citep{Adachi_et_al_1976, Greenberg_1978}, Eq. (\ref{eq:without_gas1}) and Eq. (\ref{eq:without_gas2}) can be modified as \citep{Marzari_Scholl_2000}
\begin{eqnarray}
\frac{\text{d}h}{\text{d}t} &=& Ak-B \cos{\varpi_{\rm{p}}} - Dh(h^2+k^2)^{1/2}, \label{eq:equilibrium_1} \\ 
\frac{\text{d}k}{\text{d}t} &=& -Ah + B \sin{\varpi_{\rm{p}}} - Dk(h^2+k^2)^{1/2}. \label{eq:equilibrium_2}
\end{eqnarray}

The gas drag coefficient $D$ is given by (see Eq. (4.21) and (4.14) in \citet{Adachi_et_al_1976})
\begin{eqnarray}
D &=& \frac{1}{\tau_0} \sqrt{\frac{5}{8}},\\
\tau_0 &=& \frac{2m}{\pi C_D r^2 \rho_{\rm{gas}}(a) v_{\rm{K}}(a)}.
\end{eqnarray}

Setting the left-hand side in Eq. (\ref{eq:equilibrium_1}) and (\ref{eq:equilibrium_2}) to be zero ($\text{d}h/\text{d}t = 0, \text{d}k/\text{d}t = 0$), we can solve for the equilibrium locus of $(k, h)$, i.e., the equilibrium of the eccentricity vector when the effect of secular perturbation and gas drag balance. 

At the equilibrium, the orbital eccentricity and longitude of pericenter are 
\begin{eqnarray}
e &\quad =\quad & -\frac{A}{D} \tan{\varpi} \\
\tan^2{\varpi} &\quad =\quad & \frac{1}{2} \Big\{ -1 + \Big[1+\frac{4(BD)^2}{A^4}\Big]^{1/2} \Big\}.
\end{eqnarray}


\begin{figure*}
\gridline{\fig{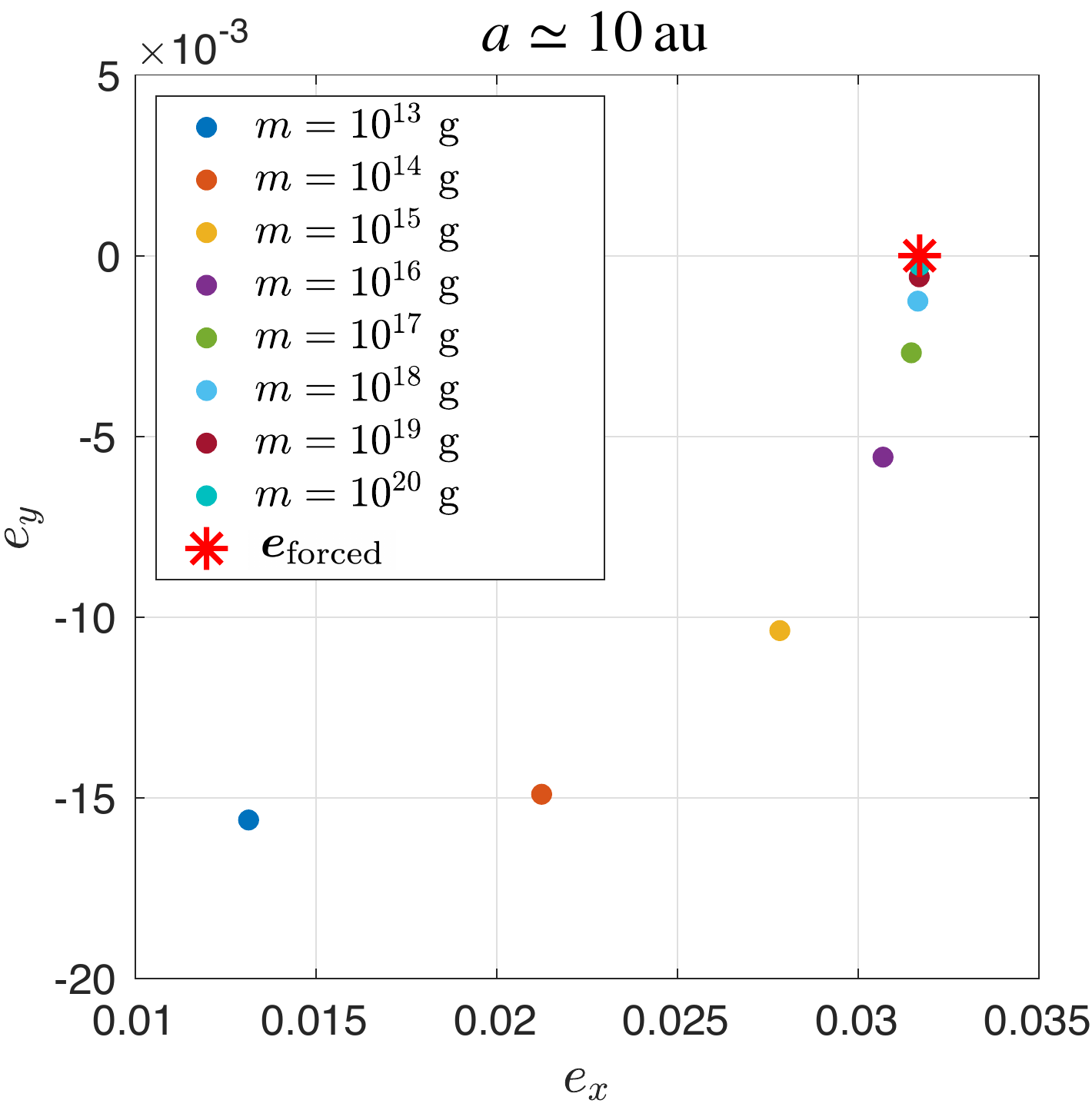}{0.343 \textwidth}{(a)}
	\fig{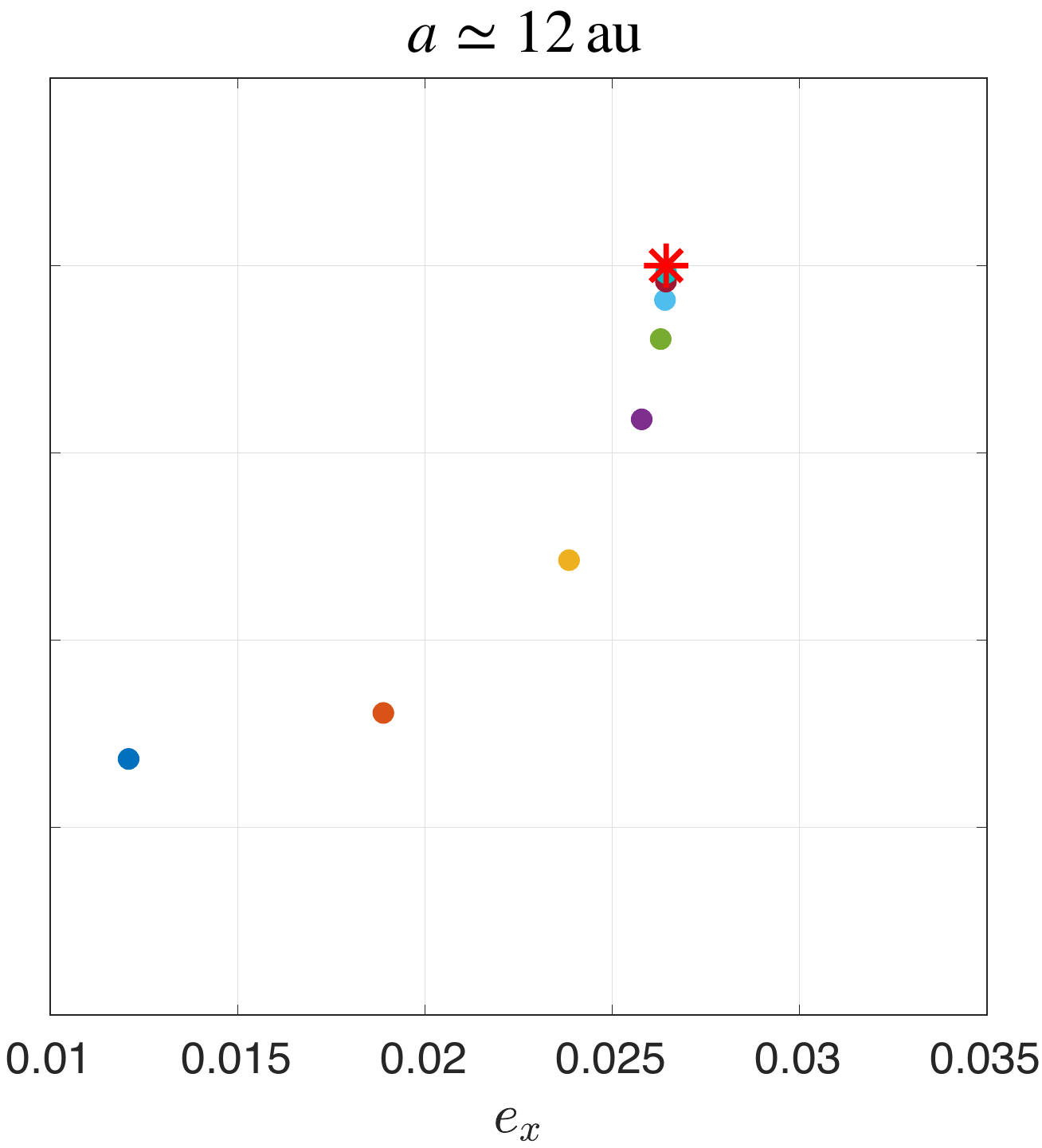}{0.318 \textwidth}{(b)}
	\fig{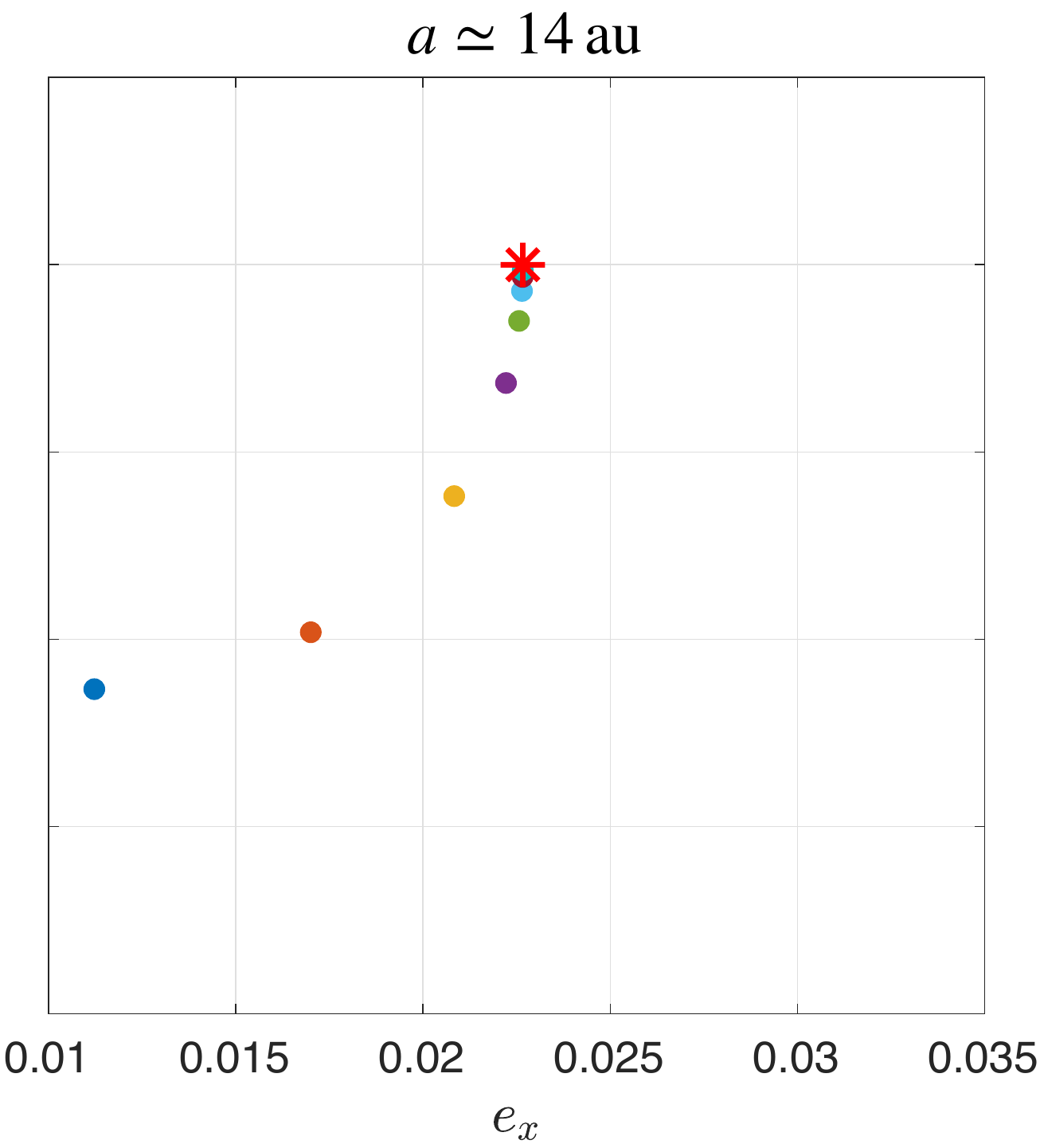}{0.318 \textwidth}{(c)}}
\caption{Equilibrium loci of pericenters on the $e_x$-$e_y$ plane near (a) 10, (b) 12, and (c) 14 au. The different symbols for data points indicate different particle masses.}
\label{fig:hk_eq}
\end{figure*}

Fig. \ref{fig:hk_eq} shows the equilibrium loci of the eccentricity vector at 10, 12, and 14 au as examples. 
Since the pericenter of the planet is on the $x$-axis ($\varpi_{\rm{p}} = 0$), the locus of the forced eccentricity (marked by the red dots) is also on the $x$-axis at any orbital distance. 
As the orbital distance from the planet increases, the forced eccentricity moves closer to the origin as a result of weaker secular perturbation. 
The equilibrium loci of different masses are marked by different symbols. 
As the particle mass increases, the equilibrium of the eccentricity vectors approaches the forced eccentricity. 
We can interpret in this way: as the particle mass increases, the gas drag becomes weaker. In the limit of zero-gas drag ($D = 0$), the solutions to $\text{d}h/\text{d}t = 0$ and $\text{d}k/\text{d}t = 0$ are $(k, h) = (e_{\rm{forced}} \cos{\varpi_{\rm{p}}}, e_{\rm{forced}} \sin{\varpi_{\rm{p}}})$, which is the locus of the forced eccentricity.
In the presence of gas drag, the path of $(k, h)$ follows a spiral centered at $(e_{\rm{forced}} \cos{\varpi_{\rm{p}}}, e_{\rm{forced}} \sin{\varpi_{\rm{p}}})$ with decreasing radius, until it reaches the equilibrium locus, as shown in Fig. \ref{fig:hk_eq}.


\section{Results}
\label{sec:results}

\begin{figure}[ht!]
\centering
\includegraphics[scale=0.55]{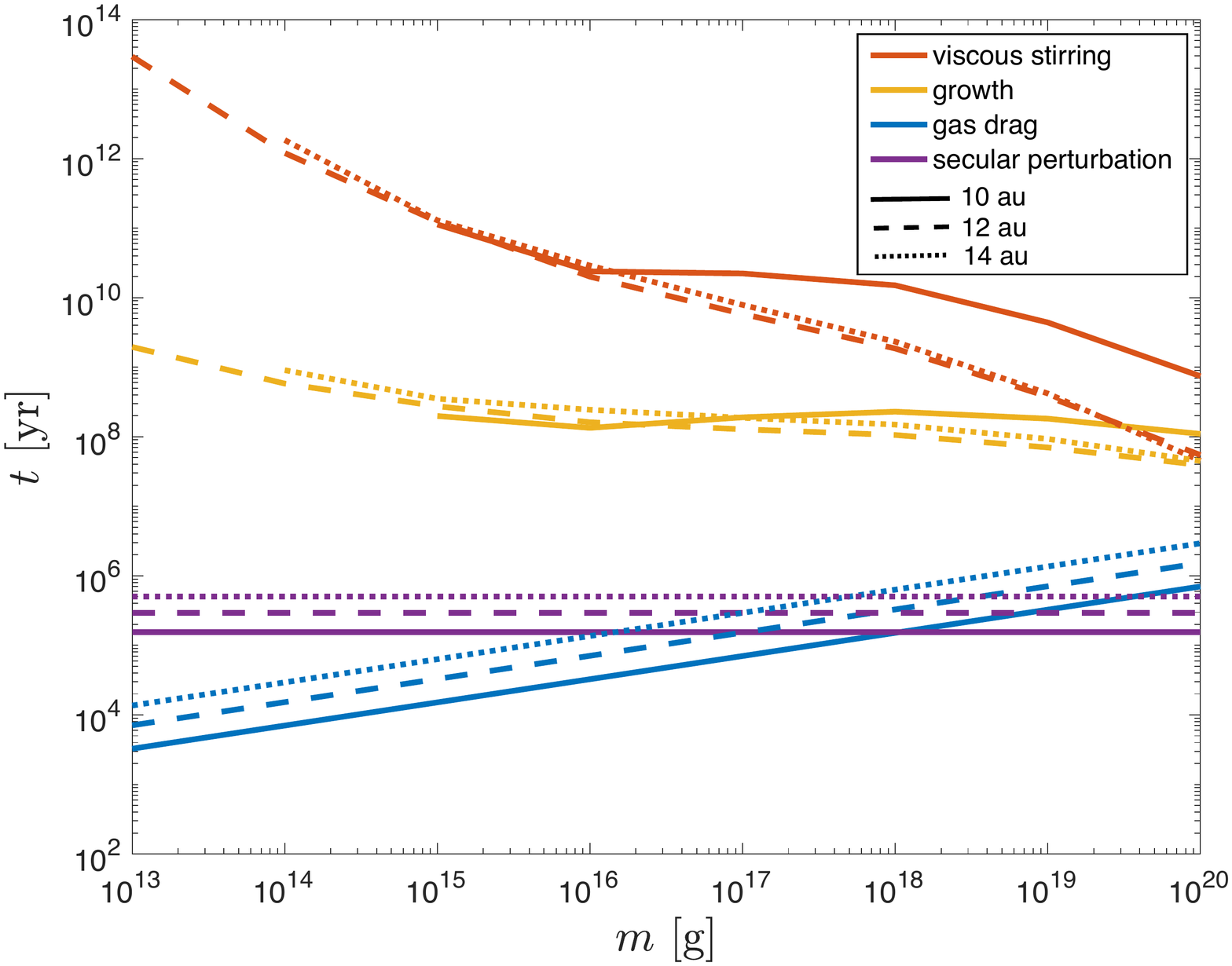}
\caption{Rough estimates of timescales for viscous stirring (red), particle growth (yellow), gas drag (blue), and secular perturbation (purple) to come into effect. The solid, dashed, and dotted lines are for timescales estimated at 10, 12, and 14 au, respectively. Some parts of the lines are missing (e.g., the red and yellow solid lines at $10^{13}$ g and $10^{14}$g, and the red and yellow dotted lines at $10^{13}$ g) due to a deficiency of particles of these masses at the given semi-major axis. This is because small particles drift inward fast due to gas drag.}
\label{fig:timescales}
\end{figure}

\subsection{Time evolution}
\label{sec:time_evolution}

\subsubsection{Timescales}
\label{sec:timescales}

To see the overall evolution, we first estimate the timescales of viscous stirring, gas drag, and secular perturbation. 
We also estimate the timescale of particle growth as a reference, although our simulation does not cover this process.
The purpose of the estimates is not to give a precise evaluation of the timescales of each physical process. 
Instead, it is to help us better understand the general evolution of the system under the influence of multiple physical processes. 

We calculate the timescales of secular perturbation and gas drag by Eq. (\ref{eq:tau_sec}) and Eq. (\ref{eq:tau_gas}) respectively. For the term $e$ in Eq. (\ref{eq:tau_gas}), we use the values of the forced eccentricity $e_{\rm{forced}}$ at different semi-major axis.

The timescale for viscous stirring $\tau_{\rm{VS}}$ is given by \citep{IDA1993210}
\begin{equation}
\tau_{\rm{VS}} = \frac{\sigma_v^4}{2 \pi G^2 m \Sigma_{\rm{d}} n \ln{\Lambda}},
\end{equation}
Here, $n$ is the Keplerian angular velocity, $\Sigma_{\rm{d}}$ is the surface density of planetesimals, and $\ln{\Lambda}$ is the Coulomb logarithm given by $\ln(b_\mathrm{max}/b_\mathrm{min})$, 
where $b_{\rm{max}}$ is the maximum impact parameters for the encountering bodies and $b_{\rm{min}}$ is the $90^{\circ}$ deflection radius \citep{binney2011galactic, IDA1990129}. 
Taking $b_{\rm{max}}$ to be equal to the vertical thickness of the planetesimal disk yields the estimate $\ln{\Lambda} \simeq 9$ \citep{armitage2010astrophysics}. 
Here we use the solid surface density $\Sigma_{\rm{d}}$ given by the minimum mass solar nebula (MMSN) as in Eq. (\ref{eq:Sigma_d}).

For the growth timescale for a planetesimal with mass $M$, we use Eq. (6) in \citet{Kokubo_Ida_2002}
\begin{equation}
\tau_{\rm{growth}} \simeq 4.8 \times 10^3   \langle \tilde{e}^2 \rangle  \left(\frac{M}{10^{26} \text{ g}}\right)^{1/3} \left(\frac{\rho_{\rm{p}}}{2 \text{ g } \rm{cm}^{-3}}\right)^{1/3}  \left(\frac{\Sigma_{\rm{d}}}{10 \text{ g } \rm{cm}^{-2}}\right)^{-1} \left( \frac{a}{1 \text{ au}}\right)^{1/2} \left(\frac{M_*}{M_{\odot}}\right)^{-1/6} \text{ yr},
\label{eq:growth_timescale}
\end{equation}
where $\langle \tilde{e}^2 \rangle ^{1/2}$ is the reduced rms eccentricity given by $\langle \tilde{e}^2 \rangle ^{1/2} = \sigma_e /h$, and $h$ is the reduced Hill radius of a growing body of mass $M$ defined as $h = \left( \frac{M}{3M_*}\right)^{1/3}$,
and $\rho_{\rm{p}}$ is the material density of planetesimals, for which 1 g $\rm{cm}^{-3}$ is used here. 
The values of $\sigma_e$ used in the calculation of the viscous stirring timescale and growth timescale are taken from simulation data and are listed in Table \ref{tab:sigma_e}. 

\begin{deluxetable*}{C | C C C}
\tablenum{2}
\tablecaption{eccentricity dispersion $\sigma_e$ \label{tab:sigma_e}}
\tablewidth{0pt}
\tablehead{\colhead{ } & \colhead{10 au} & \colhead{12 au} & \colhead{14 au}}
\startdata
10^{13} \rm{g} & \rm{-} & 0.0021 & \rm{-} \\
10^{14} \rm{g} & \rm{-} & 0.0017 & 0.0017 \\
10^{15} \rm{g} & 0.0017 & 0.0017 & 0.0016 \\
10^{16} \rm{g} & 0.0021 & 0.0018 & 0.0020 \\
10^{17} \rm{g} & 0.0034 & 0.0024 & 0.0026 \\
10^{18} \rm{g} & 0.0049 & 0.0030 & 0.0032 \\
10^{19} \rm{g} & 0.0068 & 0.0035 & 0.0038 \\ 
10^{20} \rm{g} & 0.0075 & 0.0039 & 0.0038 \\
 \enddata
\tablecomments{This table shows the values of $\sigma_e$ used in calculating the viscous stirring and growth timescales. The data for 10 au are taken from snapshots at $t = 210004$ yr; similarly, we use snapshots at $t = 300007$ yr for data for 12 au, and $t = 510012$ yr for data for 14 au. The missing values ("-") are due to a shortage of data caused by strong gas drag and fast inward drift of low-mass particles. All values are accurate to four decimal places.}
\label{tab:sigma_e}
\end{deluxetable*}

Fig. \ref{fig:timescales} shows the results for the timescales at 10, 12, and 14 au as functions of particle mass. 
The reason why we choose these three locations will be explained in the following sections.
Since the results for viscous stirring and particle growth are calculated from simulation data, there are missing data points at certain masses due to inward drifting and resonance trapping. 
For example, at 10 au, the blue and red solid lines are missing at particle masses of $10^{13}$ and $10^{14}$ g, because there are no such particles in this range of semi-major axis at the time of retrieving the data. 
This is due to fast inward drift of small particles by gas drag. 
However, at 12 au, the dashed lines are all complete. 
This is because the particles are trapped from drifting inward by mean motion resonance (MMR) near 11 au (3:1).
The times at which we retrieve data ($\sigma_e$) for calculating the viscous stirring and growth timescale are determined by the secular perturbation timescales (the purple lines), by which the eccentricity vectors of the planetesimals have completed one round of circulation around the forced eccentricity. 
This is not a very precise choice - $\sigma_e$ does not saturate or reach equilibrium for all particle masses at the secular perturbation timescale. 
However, the drastic increase in $\sigma_e$ due to relaxation of the initial conditions (low $\sigma_e$ at $t=0\,$yr) has already completed before the secular perturbation timescale. 
Afterwards, $\sigma_e$ slowly and steadily increases for several particle masses under the influence of the planet's perturbation as well as gas drag; for small particle mass like $m=10^{13}\,$g, $\sigma_e$ has already saturated by the secular perturbation timescale.
Therefore, we adopt the $\sigma_e$ values at the secular perturbation timescale to calculate the timescales of viscous stirring and growth for a qualitative estimate.
Here we use $t = 210004$ yr for 10 au, $t = 300007$ yr for 12 au, and $t = 510012$ yr for 14 au.

The secular perturbation timescale $\tau_{\rm{sec}}$ does not depend on particle mass, and it slightly increases as $a$ increases. 
According to the interval of data output in our simulations, we pick snapshots at $t = 525012$ yr (the nearest time to $\tau_{\rm{sec}}$ at 14 au) to display the pericenter alignment in later sections. 
For most of the particle masses, except for the largest three masses $m=10^{18}$, $10^{19}$, and $10^{20}$ g, the timescale of gas drag $\tau_{\rm{gas}}$ is smaller or comparable to that of secular perturbation $\tau_{\rm{sec}}$. 
For the largest masses, the gas drag timescale is $\sim \mathcal{O}(1)$ Myr. 
Therefore, to visualize the effect of gas drag on the largest particles, we show results of encounter velocities at both $t \simeq \tau_{\rm{sec}}$ and 1 Myr in Section \ref{sec:v_rel}. 
For all the particle masses in our models, the timescale of viscous stirring is orders of magnitude larger than the gas drag and secular perturbation timescales, meaning that our assumption of test particles is reasonable. 
However, judging from the decreasing trend of the viscous stirring timescale and increasing gas drag timescale with particle mass, for particles of $m \gtrsim 10^{20}$ g, the results from simulations should be treated with caution. 

\begin{figure*}[h!]
\centering
\includegraphics[scale=0.5]{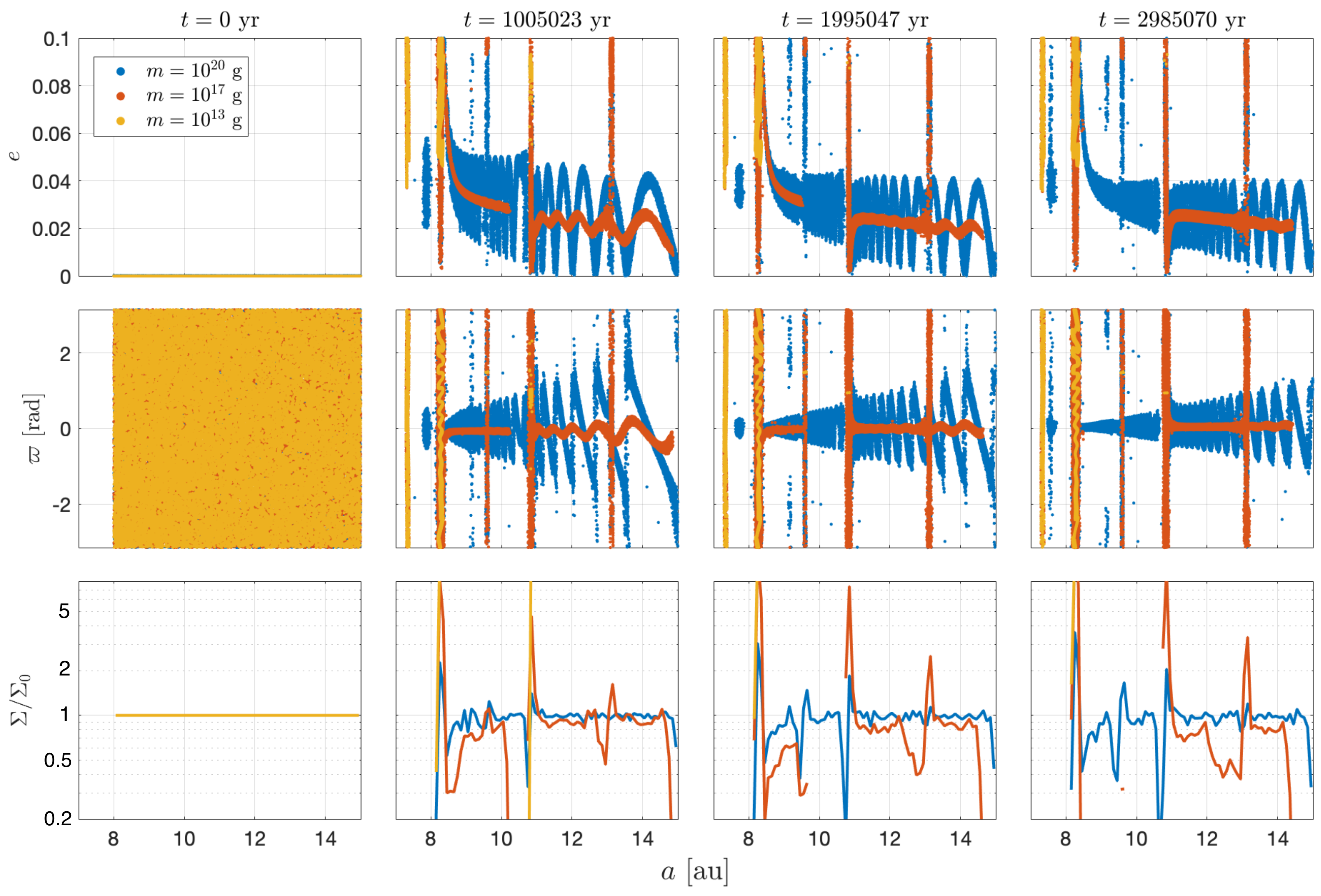}
\caption{Time evolution of the distribution of eccentricities (upper row), longitudes of pericenter (middle row), and surface density (bottom row) along the semi-major axis. Three masses are shown here as examples, representing small-mass particles, medium-mass particles, and large-mass particles. The snapshots are taken at approximately 0, 1, 2, and 3 Myr. The blue and orange dots are not easily seen in the snapshots taken at $t = 0$ yr because they are overlapped by the yellow dots. All the particles, regardless of their masses, are uniformly distributed on the $a$-$\varpi$ plane within the given range initially. $\Sigma_0$ is the initial surface density. The pile-up of particles at about 8, 11, and 13 au correspond to the 2:1, 3:1, and 4:1 MMRs.}
\label{fig:time_evolution}
\end{figure*}

For small particles, which feel strong gas drag, the time required for their eccentricity vectors to reach the equilibrium locus is short, so their orbits become aligned quickly in a timescale shorter than $10^5$ yr (see Fig. \ref{fig:timescales}), and thus exhibit low encounter velocities at early times (see the darker color on the bottom left part of the diagonals of the upper panels of Fig. \ref{fig:v_rel}, indicating lower encounter velocities). For more massive particles, it takes longer to reach equilibrium while their loci of $(k, h)$ keep circulating the forced eccentricity with decreasing radius (i.e., the proper eccentricity $e_{\rm{prop}}$ shrinks). The pericenters of large particles (say $m \gtrsim 10^{17}$ g) will become very closely located on the $e_x$-$e_y$ plane given sufficient time (longer than the gas drag timescale of these masses), resulting in well-aligned orbits and low relative velocities between particles of these masses. Since we only did a very crude estimate of the timescales in our simulations, we did not cover a sufficient timescale for the locus of pericenters of the largest particles to reach their equilibrium status. However, since our gas profile does not evolve with time, it is not of great significance or accuracy to simulate the system on a timescale comparable or longer than the gas disk lifetime ($\sim$ a few million years).

\subsubsection{System evolution}
\label{sec:system_evolution}

We next determine the overall evolution of eccentricity, longitude of pericenter, and surface density on million-year timescales to give a general idea of how the system evolves. 
Fig. \ref{fig:time_evolution} shows snapshots taken at four different times: $t = 0$ and $t \simeq 1$,  2, and 3 Myr. 
For the clarity of the plots, among the eight masses in total, we pick three masses, $10^{13}$, $10^{17}$, and $10^{20}$ g, as examples representing the smallest-mass, the medium-mass, and the largest-mass particles.
As a result of gas drag, particles (especially those with low masses) drift inward, until they become shepherded by MMR (locations specified in figure caption).  
The inward drift and pile-up is more prominent for lower-mass particles.
The overall trend of the time evolution of the eccentricity distribution is that over time the eccentricities get excited by the perturbation from the planet, especially at the MMR locations (the strip-like structures in the plots). 
Elsewhere, the distribution of the eccentricity oscillates with decreasing amplitude and period as a result of the coupled effect of secular perturbation and gas damping. 
Meanwhile, the longitudes of pericenter start from a uniform distribution in $[0, 2\pi]$, and gradually become aligned with different phase angles for different masses (except for the MMR locations), as they react to secular perturbation together with gas drag. 
This process was also explained in the analytical discussion of the eccentricity vector in Section \ref{sec:basic_dynamics}.

From these distributions, we can identify a few typical locations where the planetesimal orbits are relatively well aligned. 
For this, we avoid the MMR locations and their vicinities, where the distributions are rather chaotic. 
We also intend to make them uniformly distributed along $a$ so that they are not chosen ad hoc. 
We pick 10, 12, and 14 au, as three typical semi-major axes when presenting our results in the following sections.

\subsection{Pericenter alignment} \label{sec:pericenter}

A crucial result is the alignment of the pericenter of particle orbits, as found in previous studies (e.g., \citet{Kortenkamp_et_al_2001}, \citet{Marzari_Scholl_2000}, \citet{Marzari_et_al_2008}).
To show the pericenter alignment, we choose data from two snapshots: $t \simeq \tau_{\rm{sec}}$ and $t \simeq 1$ Myr. 
At $\tau_{\rm{sec}}$, the eccentricity vectors have finished one round of circulation around the forced eccentricity on the $e_x$-$e_y$ plane. 
Since the planetary embryos are generally thought to have formed within the lifetime of the gaseous disk, which is on the order of a million years, we also show the data at a million-year timescale ($t \simeq 1$ Myr) to examine the effect of pericenter alignment. 
We do not show data at the end of the simulation ($t \simeq 3$ Myr) because many small particles have already drifted inward significantly so that the data would not be very explanatory for displaying pericenter alignment. 

\begin{figure*}[h!]
\gridline{\fig{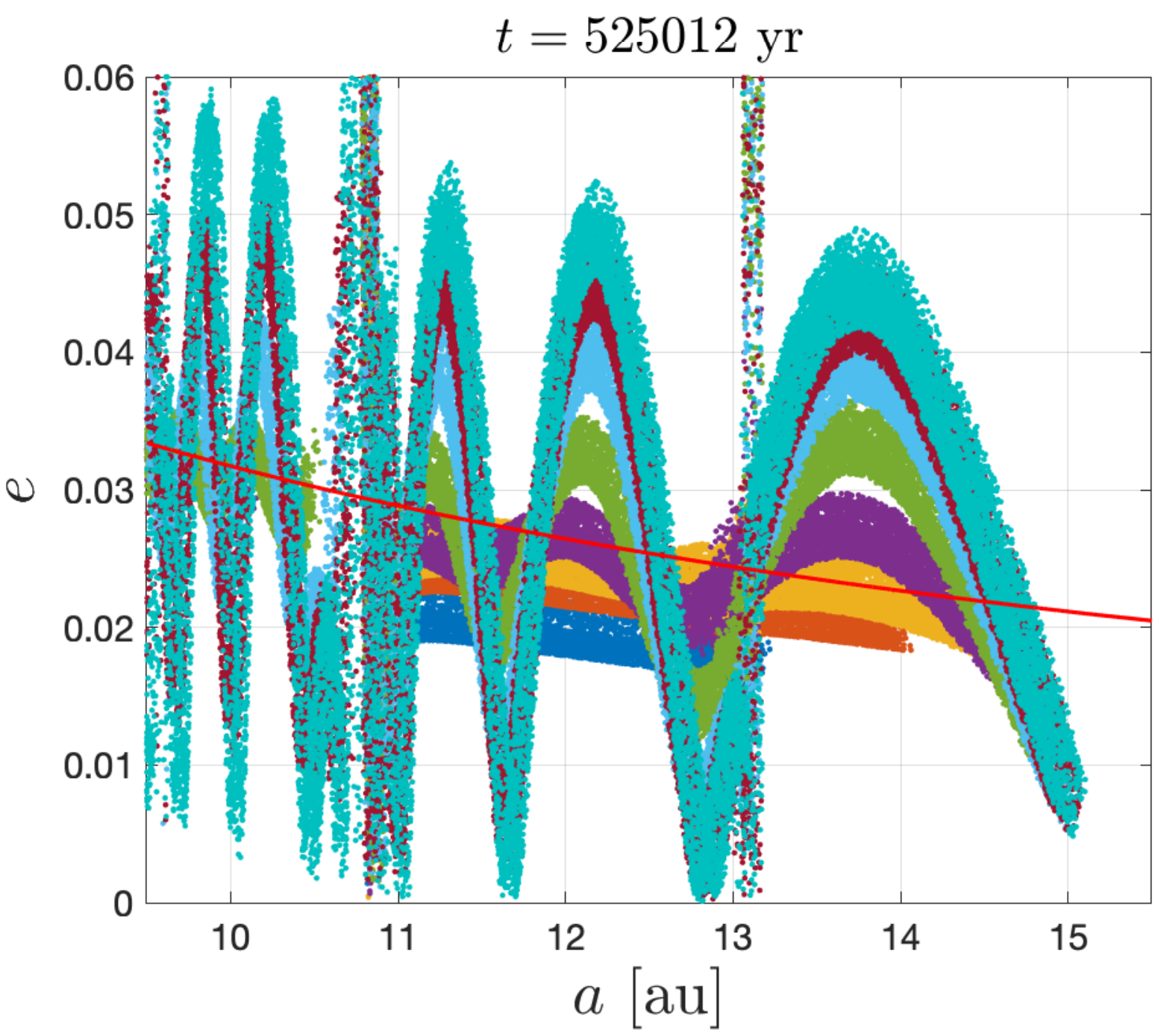}{0.49\textwidth}{(a)}
	\fig{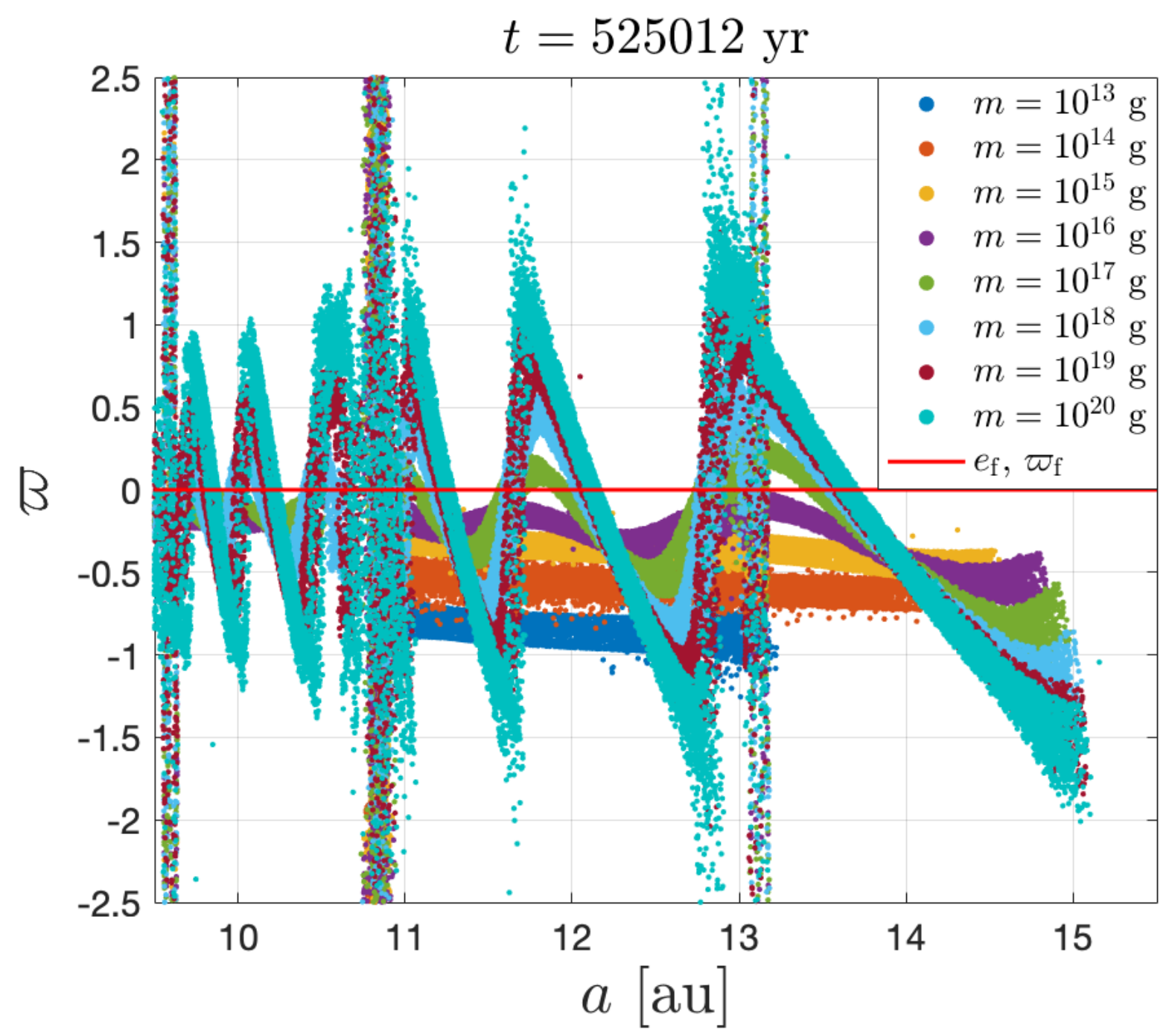}{0.5 \textwidth}{(b)}}
\caption{At $t = 525012$ yr $\simeq \tau_{\rm{sec}}$: zoomed-in views of (a) the distribution of eccentricities along the semi-major axis, (b) distribution of longitudes of pericenter along the semi-major axis. The red lines show the forced eccentricity and forced pericenter longitude respectively. Here we zoom into roughly 10-15 au to reveal the structures in this semi-major axis range.}
\label{fig:alignment_oe}
\end{figure*}

Fig. \ref{fig:alignment_oe} shows the distribution of eccentricities and longitudes of pericenter along the semi-major axis at $t = 525012$ yr (t $\simeq \tau_{\rm{sec}}$). 
For reference, we show the location of the forced eccentricity and pericenter longitude in both planes. 
At this point, the eccentricities are excited at the MMR (see Fig. \ref{fig:v_disp} for the locations of MMR) and the pericenters have already started to become aligned near 10, 12, and 14 au. 
In general, the data points oscillate around the forced value ($e_{\rm{forced}}$ and $\varpi_{\rm{f}}$) on both planes. Since the planet's longitude of pericenter is set to be 0, here $\varpi_{\rm{f}}$ is also 0. 
The smaller particles are better aligned (flatter distribution). This mass dependence will be further explained in Section \ref{sec:pericenter_m}.

\begin{figure*}
\gridline{\fig{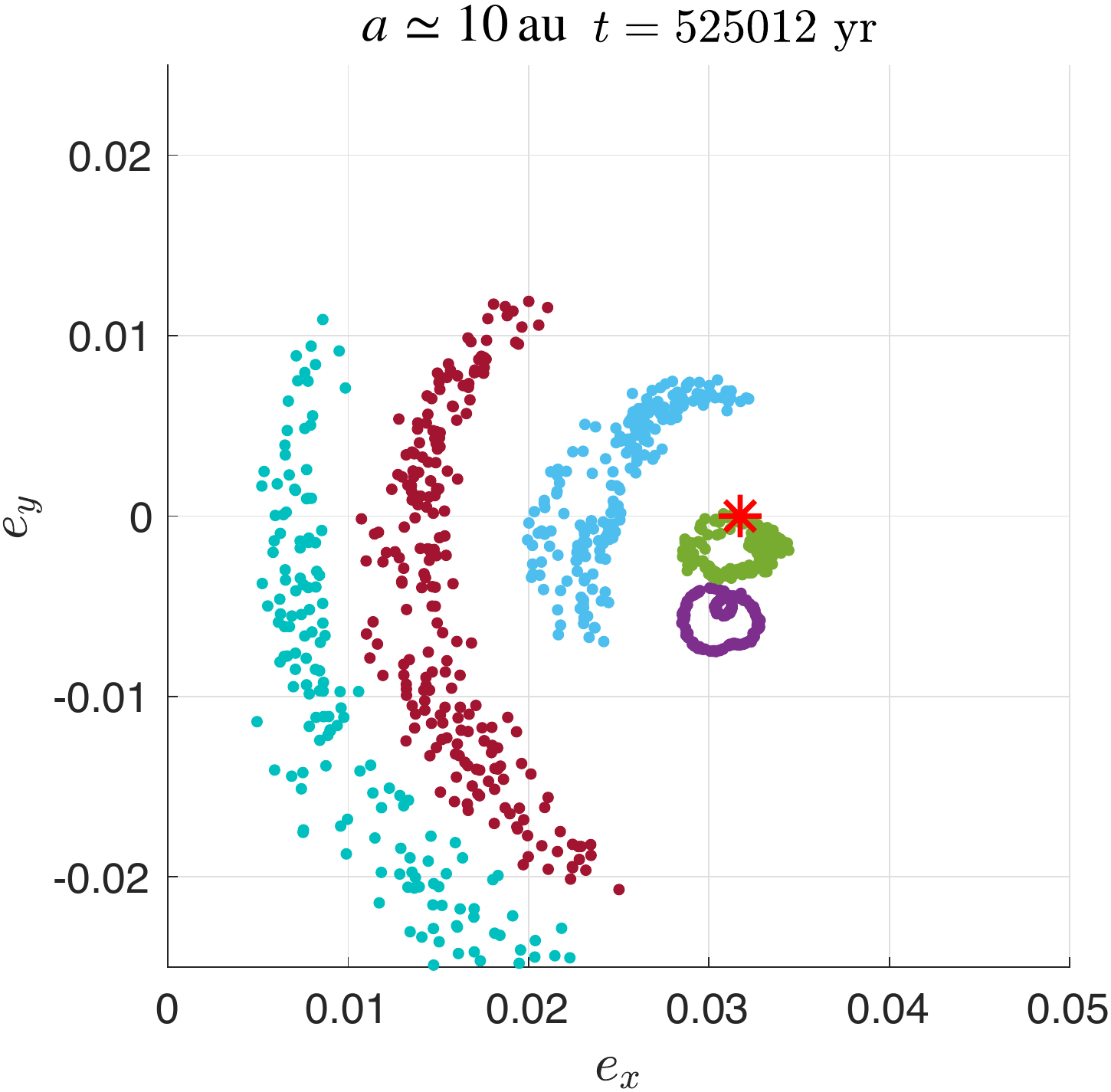}{0.3 \textwidth}{(a)}
	\fig{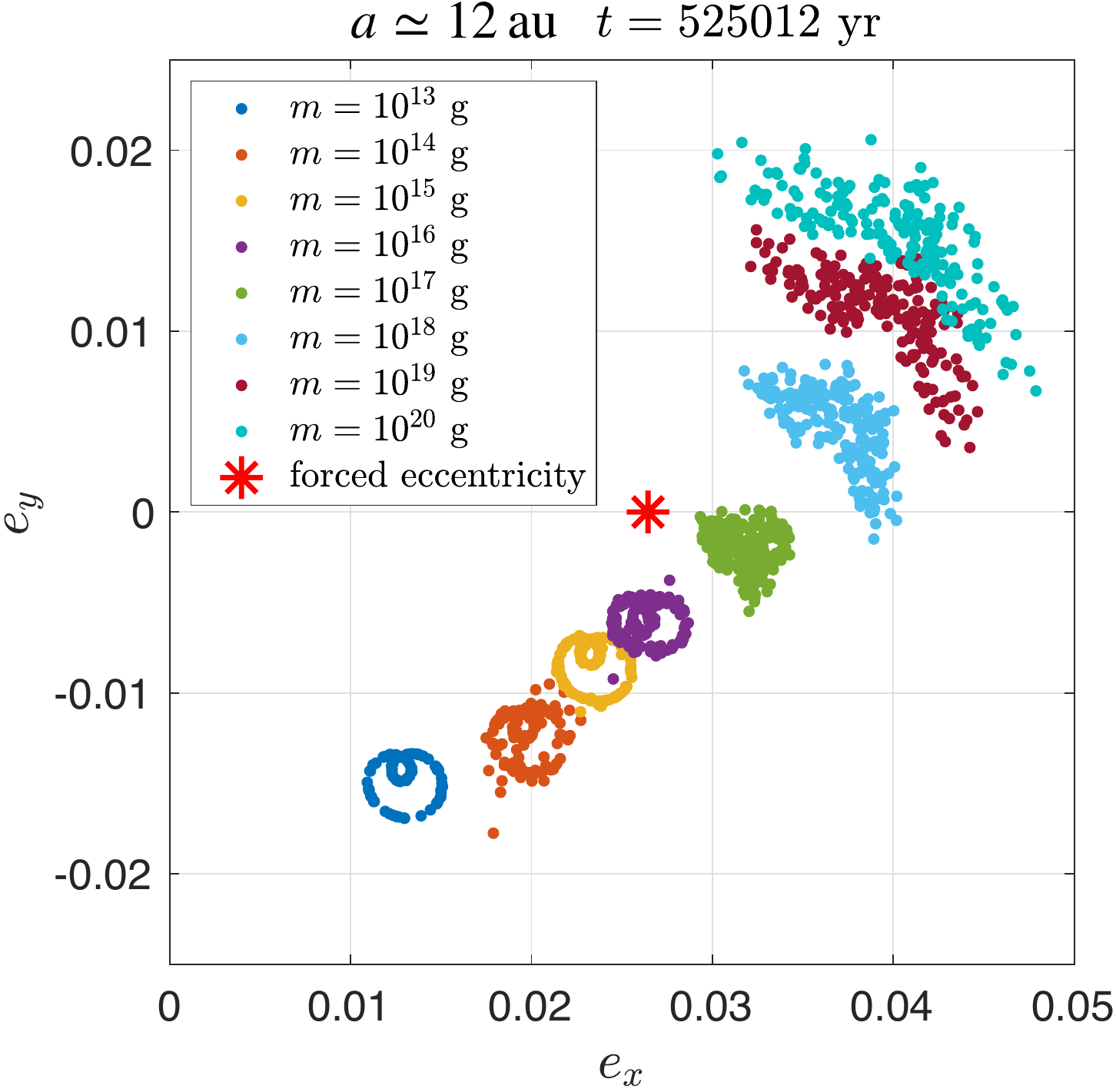}{0.3 \textwidth}{(b)}
	\fig{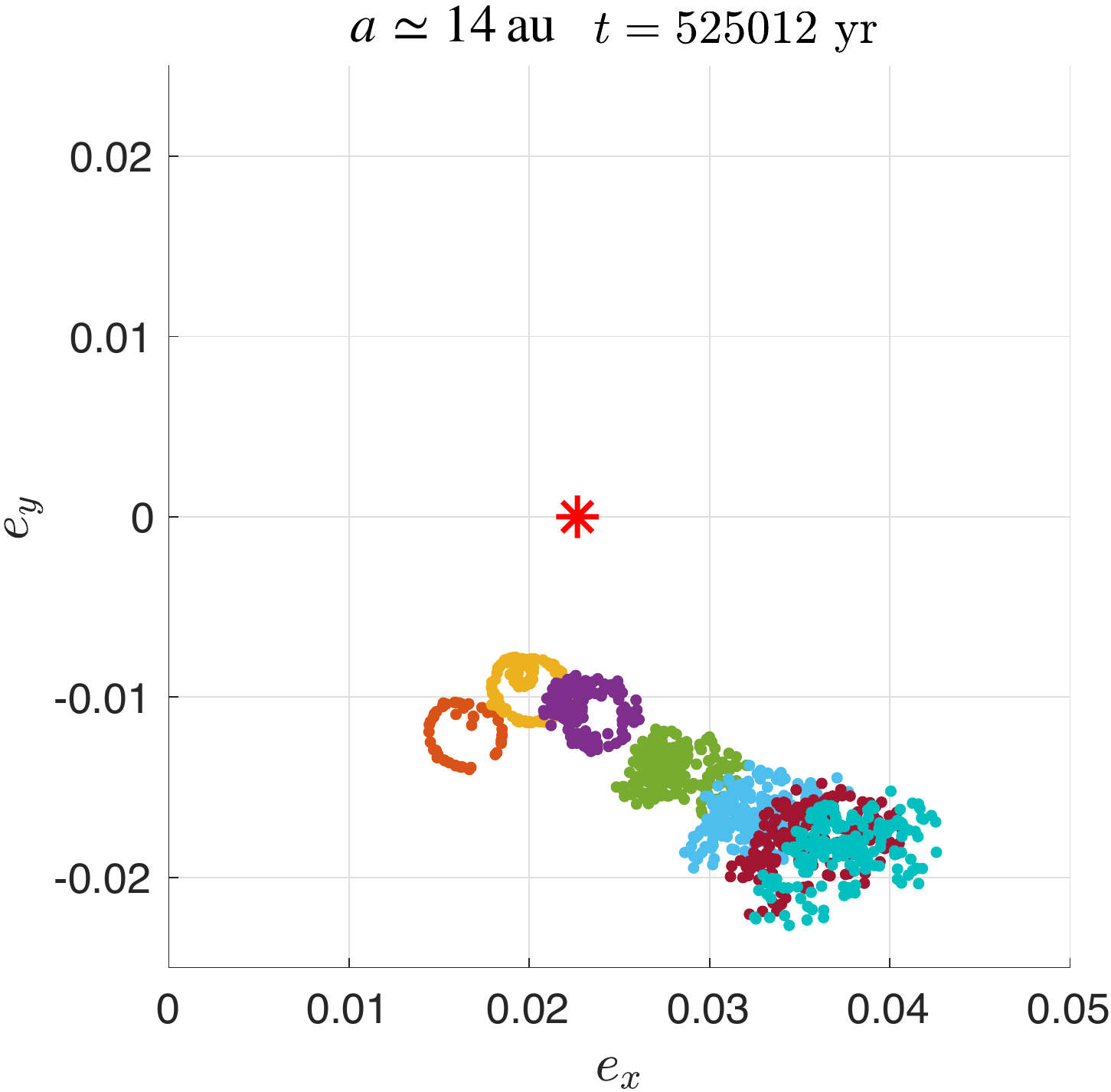}{0.3 \textwidth}{(c)}}
\gridline{\fig{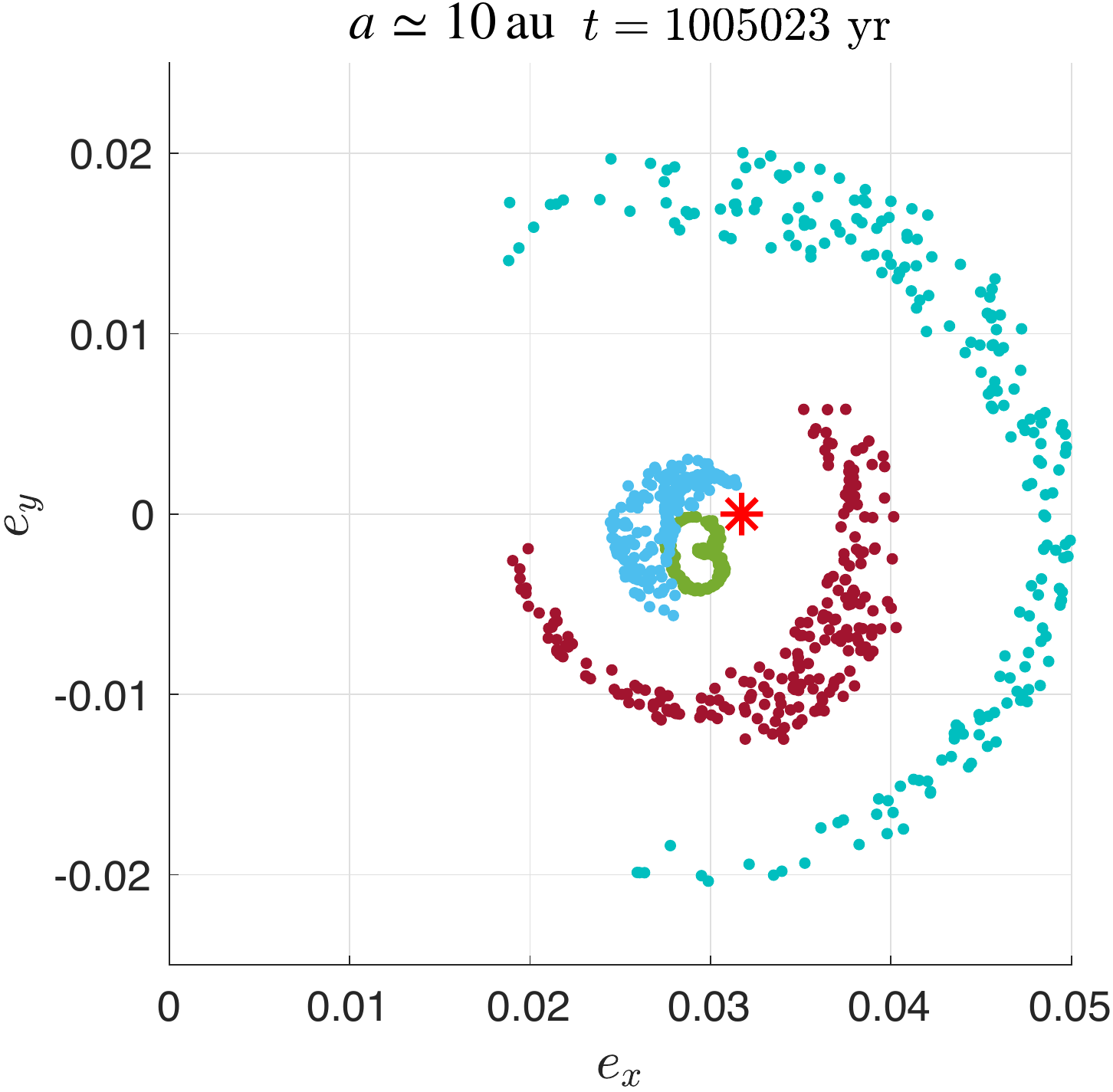}{0.3 \textwidth}{(d)}
	\fig{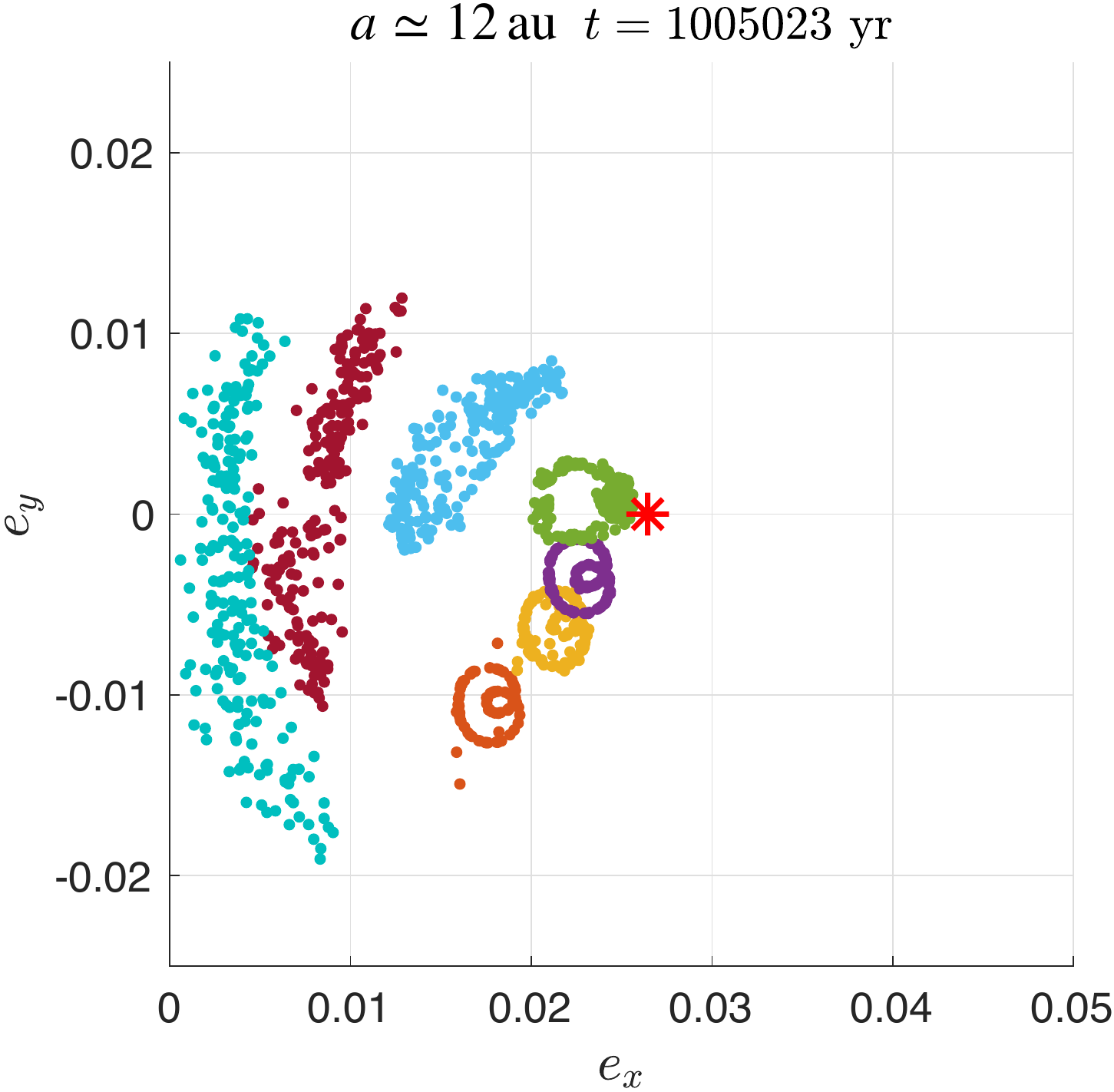}{0.3 \textwidth}{(e)}
	\fig{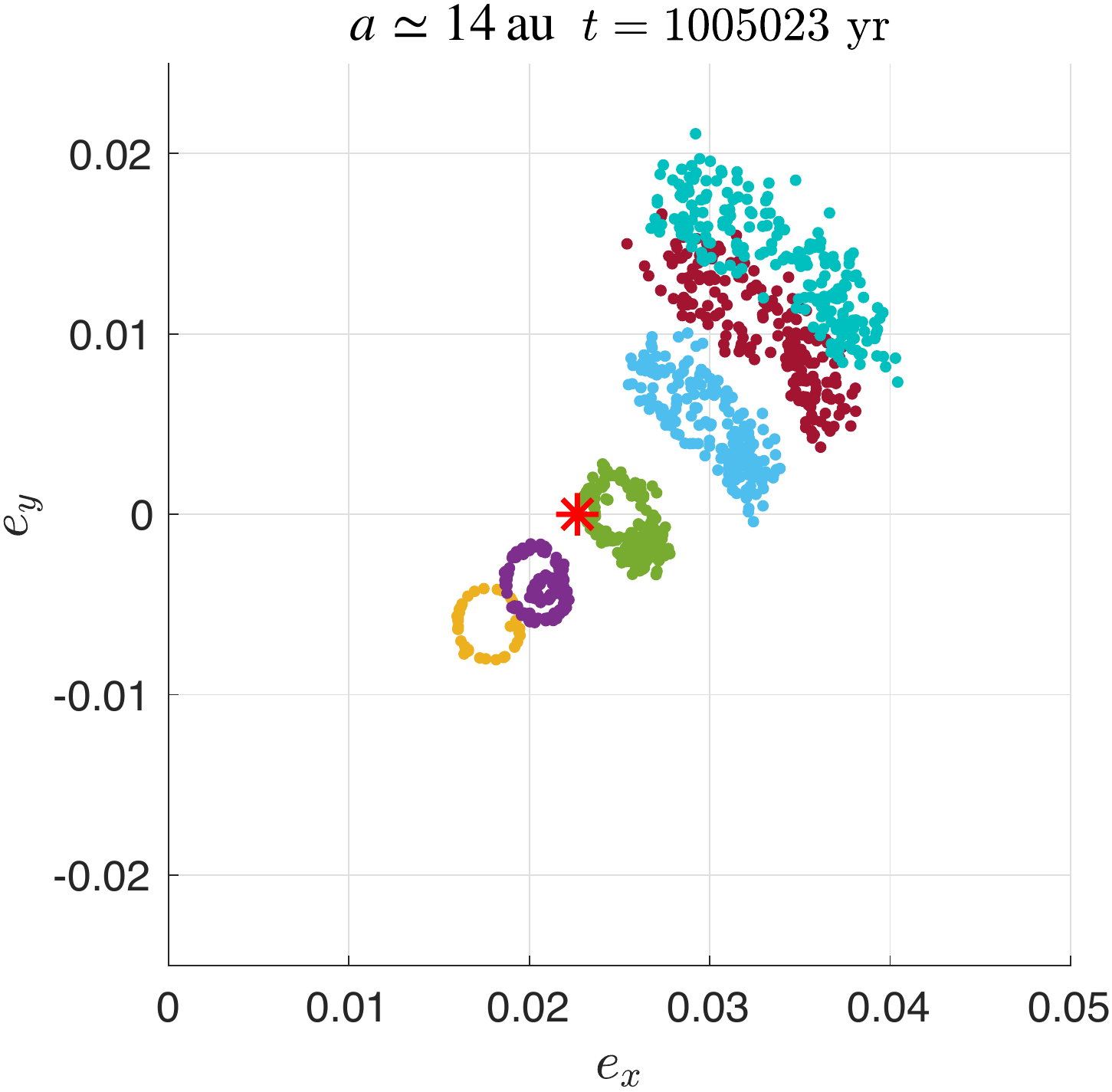}{0.3 \textwidth}{(f)}}
\caption{Distribution of eccentricity vectors (pericenters) on the $e_x$-$e_y$ plane at 10 (left), 12 (center), and 14 au (right). The first row is for snapshots taken at $t \simeq \tau_{\rm{sec}}$ (t = 505348 yr); the second row is for snapshots taken at $t = 1000670$ yr. The missing particle masses (typically on the smaller end) are due to fast orbital decay.}
\label{fig:ecc_vec_mid}
\end{figure*}

Fig. \ref{fig:ecc_vec_mid} shows the distribution of the eccentricity vectors on the $e_x$-$e_y$ plane at 10, 12, and 14 au.  
We can clearly see how the pericenters of different mass particles cluster as a result of pericenter alignment. 
The small loop structures of some clusters are due to short-term perturbations from the planet. 
The forced eccentricity is also plotted here to help visualize perturbation from the planet. 
The equation for calculating the forced eccentricity is given in Section \ref{sec:basic_dynamics}. 

We see no significant discrepancies between the two sets of data taken at the two times, except for the fact that at $t \simeq 1$ Myr, the clusters are located further from each other, indicating larger $\Delta e$ and thus larger $v_{\rm{rel}}$ between different masses. 
This is due to the fact that as time passes, the effect of gas drag gradually becomes prominent. 
Since the timescale of gas drag varies for different particle masses, the data points initially assembled near the origin are slowly pulled apart on the $e_x$-$e_y$ plane. 
This change is especially prominent for large-mass particles and it becomes easier to see at later times, since the gas drag timescales for these particles are rather long. 
For smaller-mass particles ($m \lesssim 10^{17}$ g), the effect of gas drag saturates before the secular perturbation timescale so we see no significant differences in $\Delta e$ in these two snapshots.
As for the dispersion of the identical-mass particles, the clusters of larger masses become more dispersed and spread out (in longer arcs) at $t \simeq  1$ Myr because for these large masses, the secular perturbation timescale (differential precession of $\boldsymbol{e}$) is shorter than the gas drag timescale. 
In other words, the particles first feel the effect of the secular perturbation from the planet and then the effect of the gas drag forcing the alignment of their orbits.
As a result, their eccentricity vectors spread first on the $e_x$-$e_y$ plane (differential precession of the eccentricity vector), and then become concentrated eventually near the equilibrium.

To quantitatively see the difference between alignment and non-alignment of orbits, we can compare the root mean square of the eccentricity $\langle e \rangle ^{1/2} =  \sqrt{\frac{1}{N} \Sigma^N_{i=1} e_i^2}$ and the dispersion of $\sigma_e$ (given by Eq. \ref{eq:sigma_e}) for particles at locations where their orbits are aligned and not aligned.  
We take particles with $m = 10^{20}$g as an example. 
At 12 au, where their orbits are well aligned,  $\langle e \rangle ^{1/2}$  and $\sigma_e$ are 0.0432 and 0.0048 respectively. 
At 10.8 au, where the orbital distribution is rather chaotic (3:1 MMR), $\langle e \rangle ^{1/2}$  and $\sigma_e$ are 0.0357 and 0.0283. 
Apparently, when orbits are aligned, $\sigma_e$ is much smaller than $\langle e \rangle ^{1/2}$ because the eccentricity vectors (pericenters) cluster at some non-zero locus (determined by the forced eccentricity) on the $e_x$-$e_y$ plane.
When orbits are not aligned, the two values are not significantly different, since the mean of $(e_x, e_y)$ does not deviate much from the origin. 
Therefore, to assess the alignment of orbits, we use $\sigma_e$ when calculating the velocity dispersion. 

\subsubsection{Dependence on semi-major axis} \label{sec:pericenter_a}

In Fig. \ref{fig:alignment_oe} we can see that the distribution of particle eccentricities and pericenters are rather chaotic near MMR.
As $a$ increases, the eccentricity distribution becomes smoother, and the magnitudes at the peaks generally decrease as a result of weaker perturbation. 
The period of oscillation in both planes increases with $a$ (see Eq. \ref{eq:tau_sec} in the Appendix). 
The longitudes of pericenter also become more aligned further from the planet.
Fig. \ref{fig:ecc_vec_mid} shows the distribution of pericenters on the $e_x$-$e_y$ plane at 10, 12, and 14 au.
It is clear that at larger orbital distances (further from the planet), the distribution of pericenters for identical-mass particles is more concentrated, meaning that the orbits are more aligned. 
The forced eccentricity also moves closer to the origin as $a$ increases, indicating that the effect of secular perturbation from the planet is weaker. 
The distances between clusters $\Delta e$ of different masses also become slightly smaller as the orbital distance increases. 
On one hand, the weaker perturbation at larger $a$ leads to smaller forced eccentricities being induced on the particles; on the other hand, the gas drag is also weaker at regions further out in the disk, resulting in a less prominent effect differentiating the pericenter locations of different-mass particles.

\subsubsection{Dependence on particle mass} \label{sec:pericenter_m}

Gas drag reinforces pericenter alignment among identical-sized particles. 
Therefore, from Fig. \ref{fig:alignment_oe} and Fig. \ref{fig:ecc_vec_mid}, we can see that smaller particles, which are subject to stronger gas drag, have smaller eccentricities and more aligned orbits (small $\sigma_e$). 
In contrast, larger particles (of identical mass) have higher eccentricities and a sparser distribution of pericenters on the $e_x$-$e_y$ plane (large $\sigma_e$) because they are less coupled to the gas and thus less aligned.
For similar reasons, in Fig. \ref{fig:ecc_vec_mid} the clusters of particles with adjacent mass on the larger end of the mass spectrum are located closer to each other than those on the smaller end of the mass spectrum, resulting in smaller $\Delta e$ between particles of two different masses on the larger end (especially at 12 and 14 au). 
In other words, gas drag takes a much longer time to separate the pericenter locations of particles of two different masses for larger particles. 
We can also visualize this from Fig. \ref{fig:timescales}, where the gas drag timescales exceed the secular perturbation timescales for the largest-mass particles. 
This trend no longer applies after a timescale comparable to or larger than the gas drag timescale for these large particles.  

\subsection{Random encounter velocities} \label{sec:v_rel}

\subsubsection{Time evolution of velocity dispersion}

\begin{figure*}
\centering
\gridline{\fig{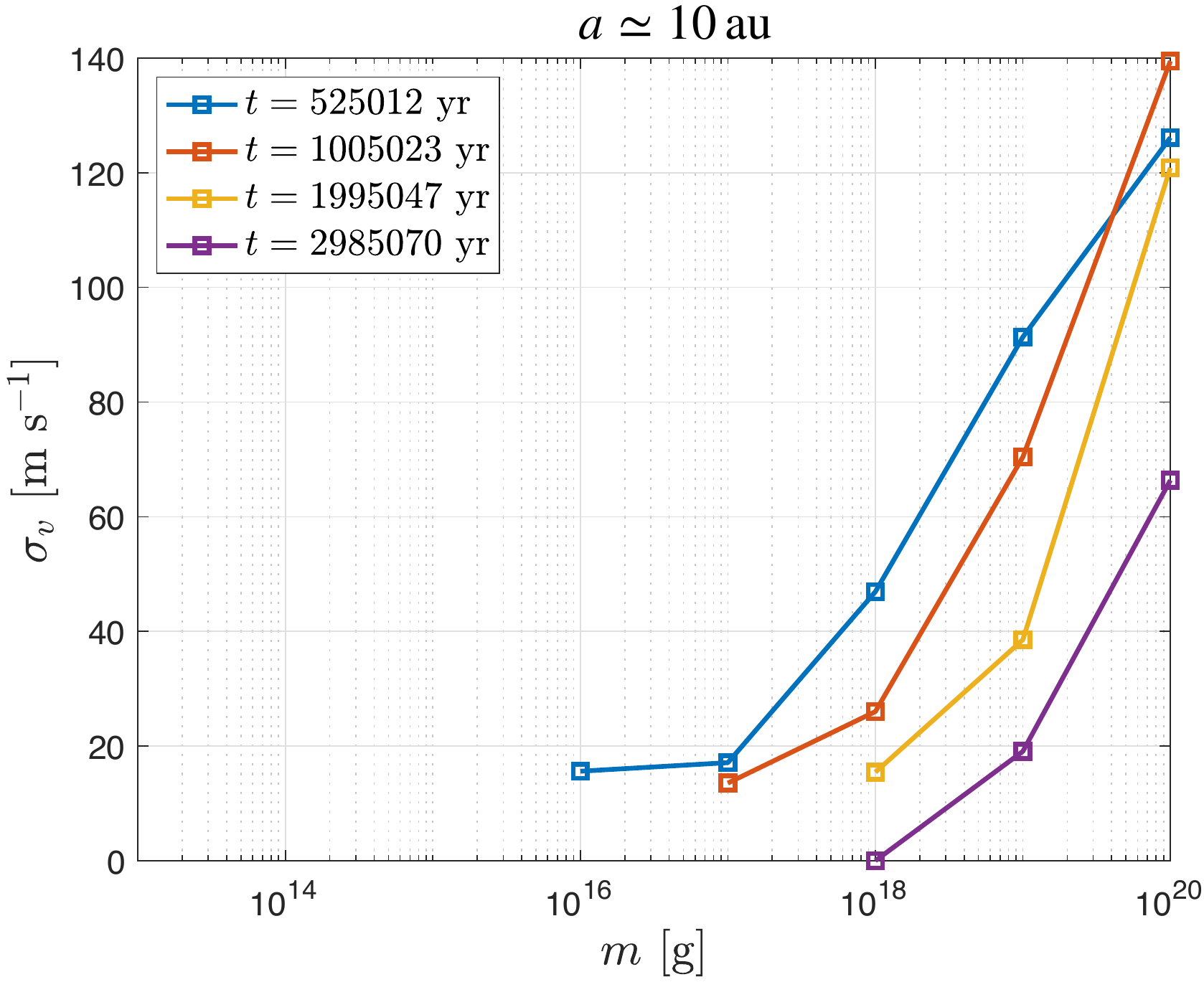}{0.34 \textwidth}{(a)}
	\fig{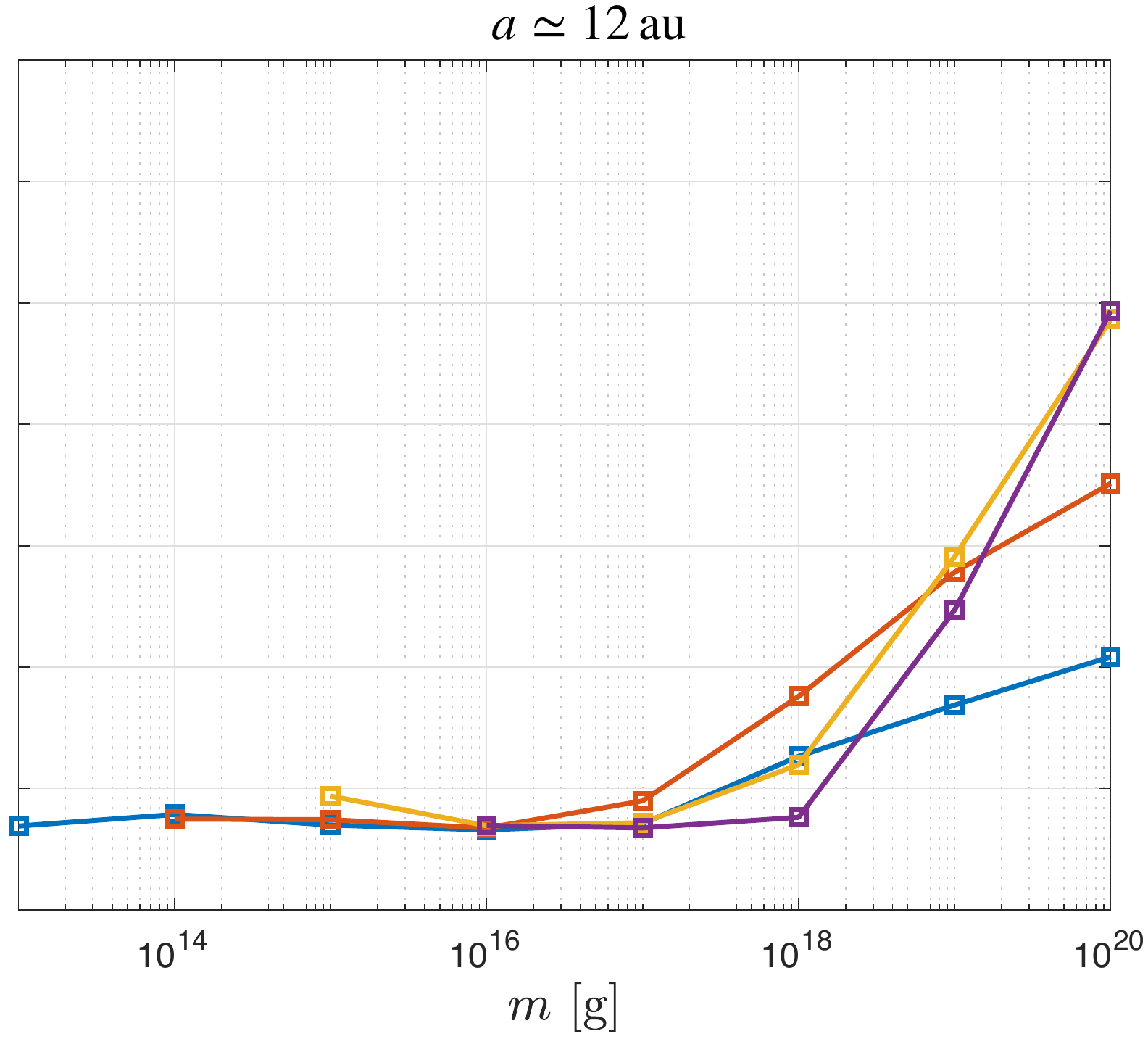}{0.308 \textwidth}{(b)}
	\fig{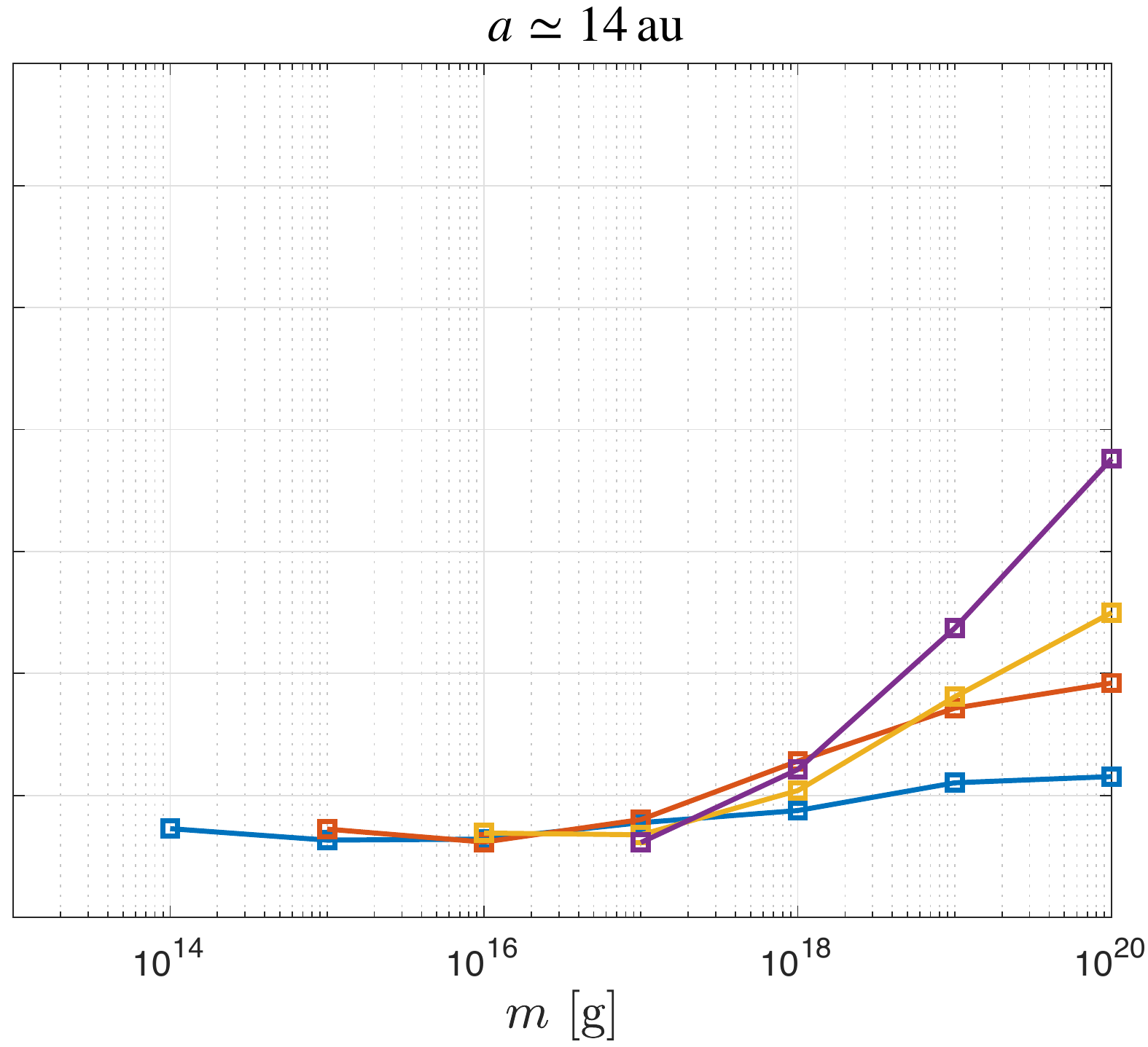}{0.307 \textwidth}{(c)}}

\caption{Velocity dispersion for different particle masses at the three given semi-major axis values. Each color shows the $\sigma_v$ values at a different time.}
\label{fig:v_disp_time}
\end{figure*}

To understand how the encounter velocities evolve with time, we determine the velocity dispersion for each mass at a given semi-major axis at four stages (Fig. \ref{fig:v_disp_time}). 
At $a \simeq 10$ au, $\sigma_v$ generally decreases monotonically over time, regardless of the particle mass. 
This is a natural result of gas damping. 
Further out at $a \simeq 12$ au, the path of the evolution becomes a little more complex. 
For $m < 18$ g, the variation in $\sigma_v$ is not significant. 
For $m \gtrsim 18$ g, $\sigma_v$ increases due to gravitational scattering and then decreases as a result of gas damping, except for the largest mass $m = 20$ g, for which $\sigma_v$  has not yet been damped at $t \simeq 3$ Myr. 
At $a \simeq 14$ au, as the gas drag becomes even weaker, $\sigma_v$ for both $m = 19$ g and $20$ g keeps increasing until $t \simeq 3$ Myr. 
This behavior was also discussed in Section \ref{sec:pericenter}, where we explained why the eccentricity vectors have larger dispersions for larger particle masses as time increases in Fig. \ref{fig:ecc_vec_mid}. 
Despite the increasing trend of $\sigma_v$ for the largest masses at the outer regions of the disk, their highest values keep decreasing as $a$ increases. 

\begin{figure*}[h!]
\gridline{\fig{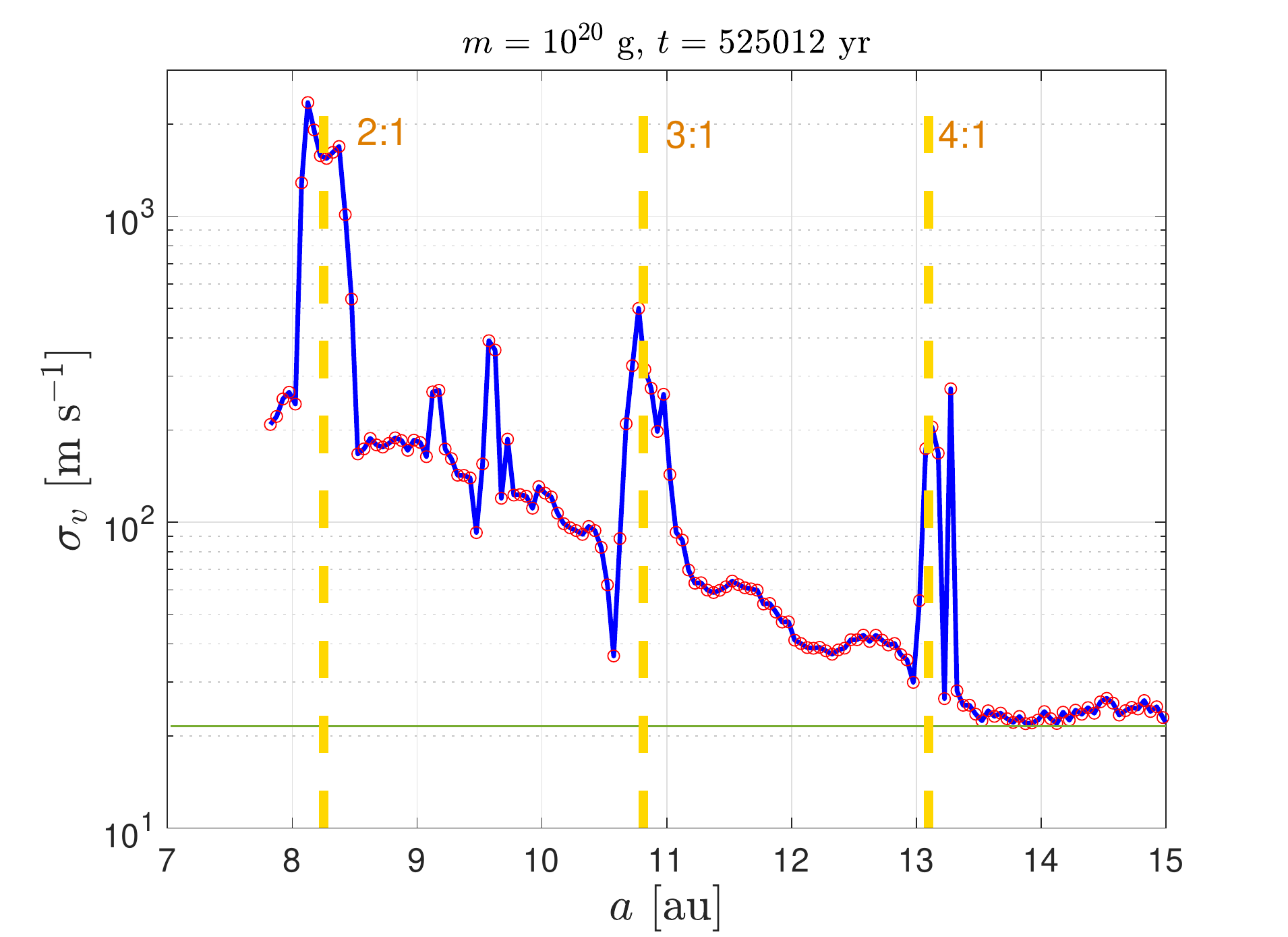}{0.47\textwidth}{(a)}
	\fig{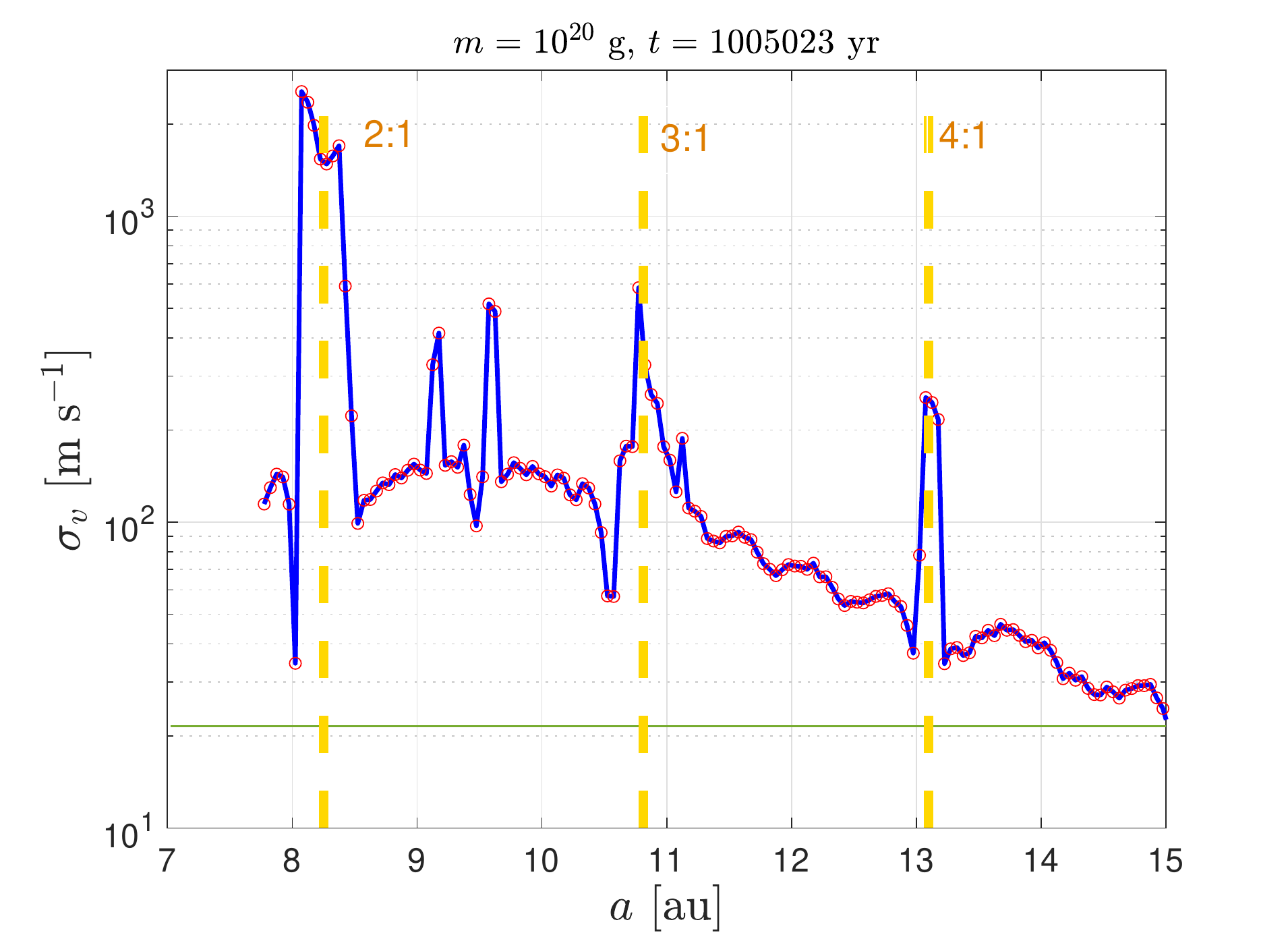}{0.47 \textwidth}{(b)}}
\caption{Distribution of velocity dispersions at (a) $t \simeq \tau_{\rm{sec}}$ and (b)  $t \simeq 1$ Myr for $m_{\rm{p}} = 10^{20}$ g. The yellow vertical dashed lines correspond to the MMR locations. The horizontal green line shows the escape velocity, given by Equation (\ref{eq:v_esc}), for reference.}
\label{fig:v_disp}
\end{figure*}

Knowing the overall time evolution, we focus on the encounter velocities at two times: $t \simeq \tau_{\rm{sec}}$ and $t \simeq 1$ Myr to check for other dependence. 
Fig. \ref{fig:v_disp} shows the velocity dispersion for $m = 10^{20}$ g along the semi-major axis as an example. 
Here we show the data of the largest particle mass because there are no missing particles due to inward drift between 8 and 15 au. 
Again, the general trend remains the same for the two snapshots: the velocities are high at the MMR locations as a result of high eccentricities and non-aligned orbits. 
Elsewhere, especially in the regions where the particle orbits are aligned, the velocities are low. 
Note that the velocity dispersion slightly increases in plot (b) due to more sufficient gravitational scattering by the planet.
Since the gas drag timescale is much longer than the secular perturbation timescale for $m = 10^{20}$ g, the velocity dispersion is not yet damped by gas drag at the time of the snapshot. 

\begin{figure*}
\gridline{\fig{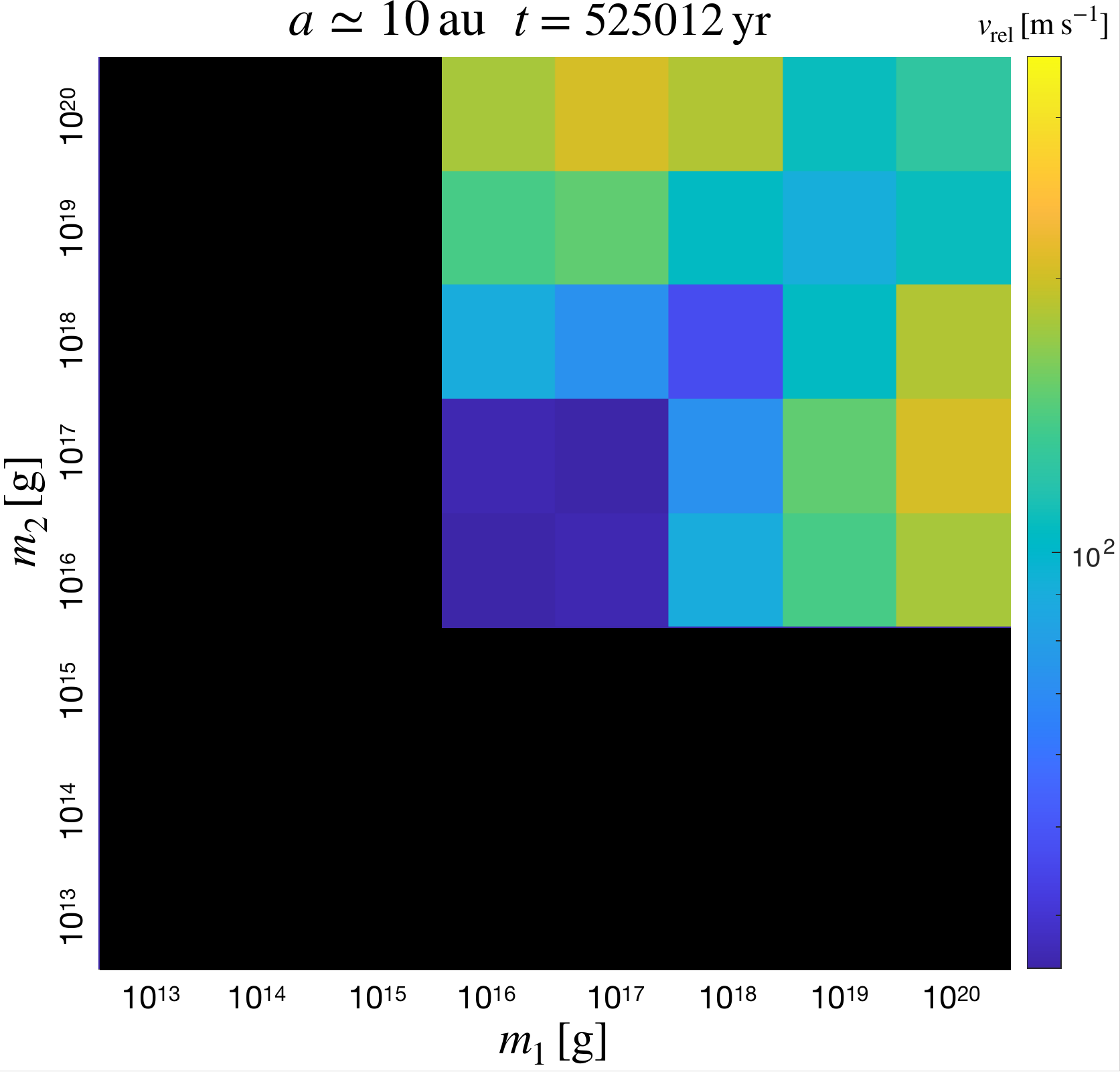}{0.3 \textwidth}{(a)}
	\fig{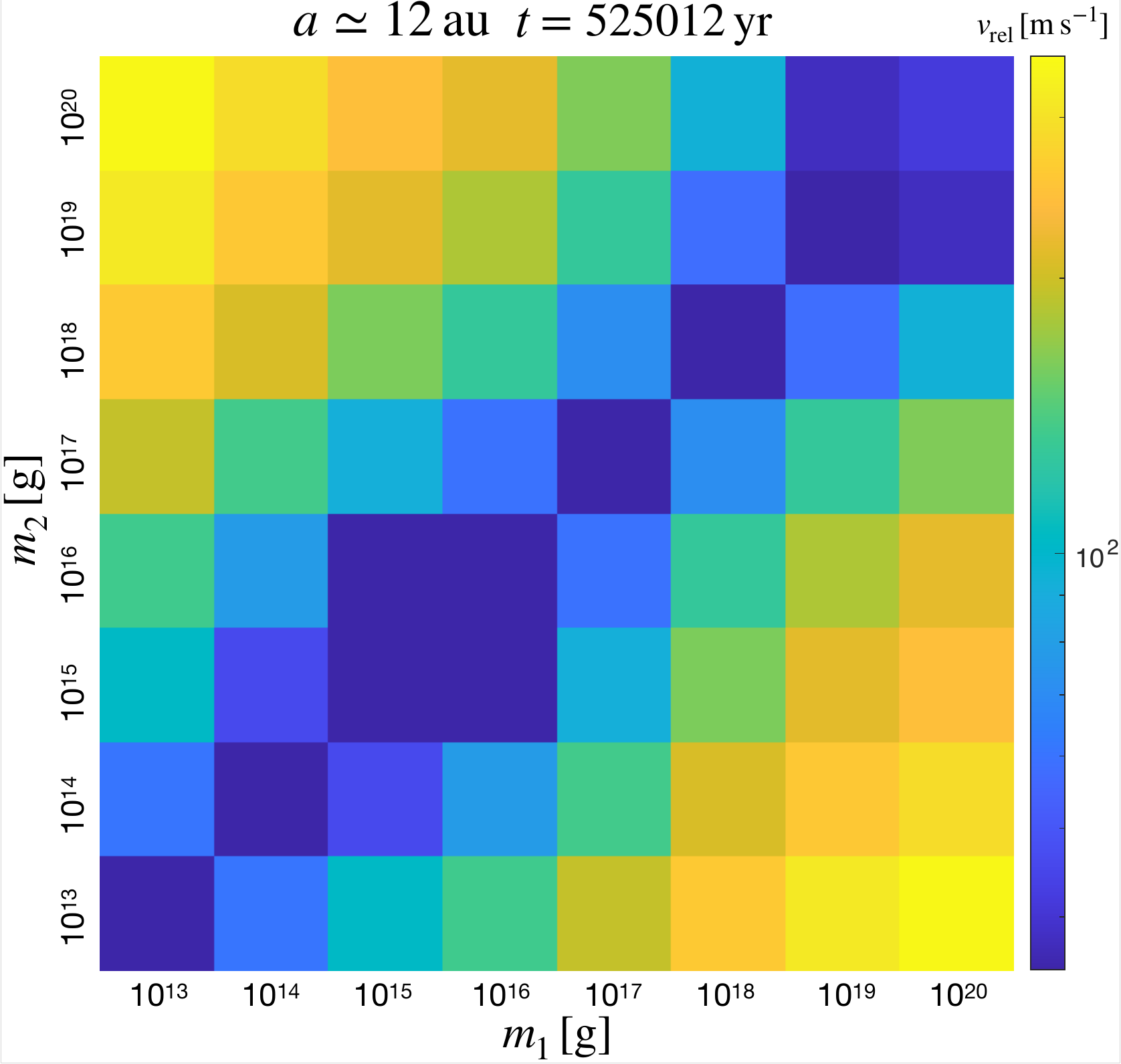}{0.3 \textwidth}{(b)}
	\fig{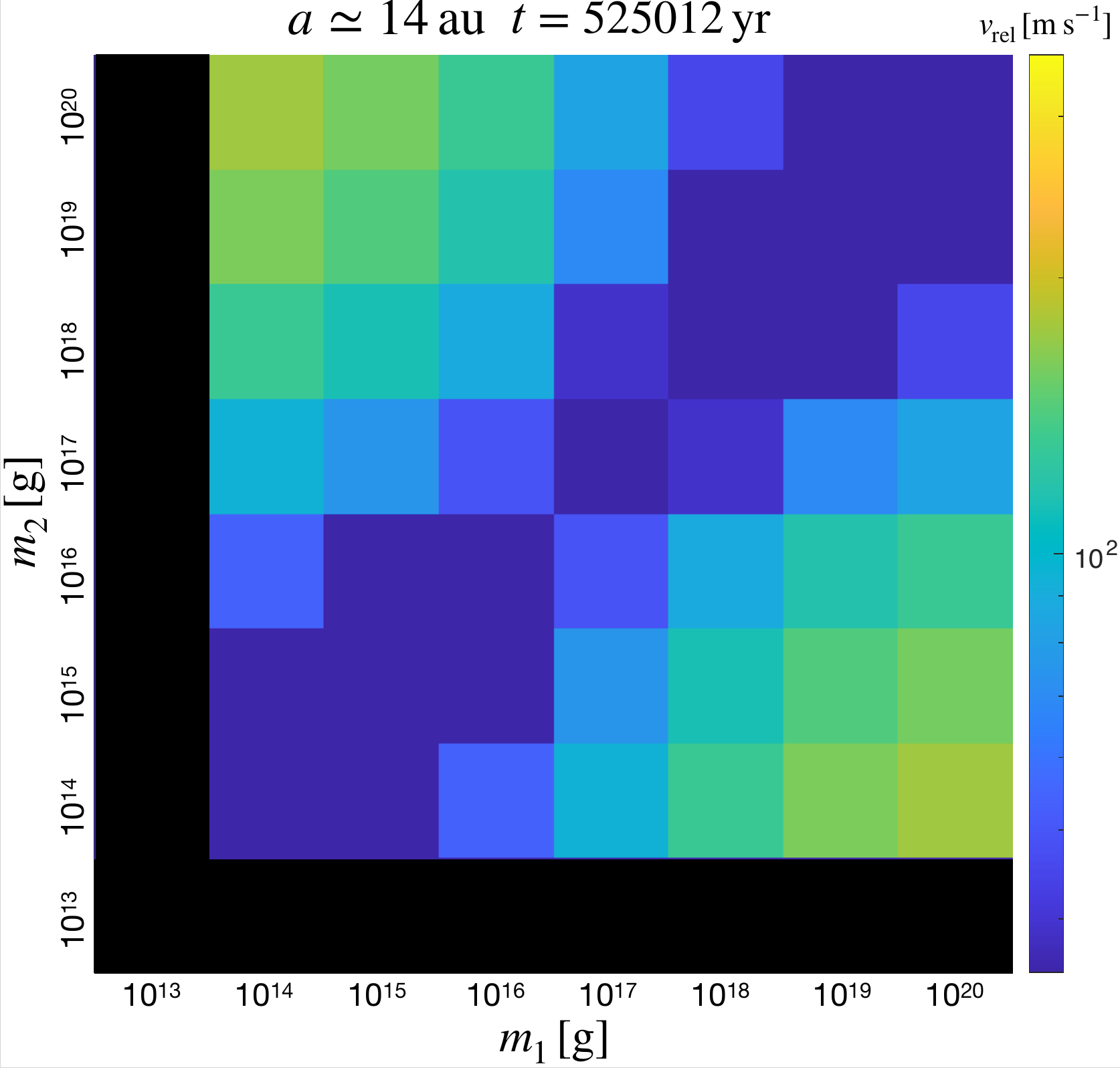}{0.3 \textwidth}{(c)}}
\gridline{\fig{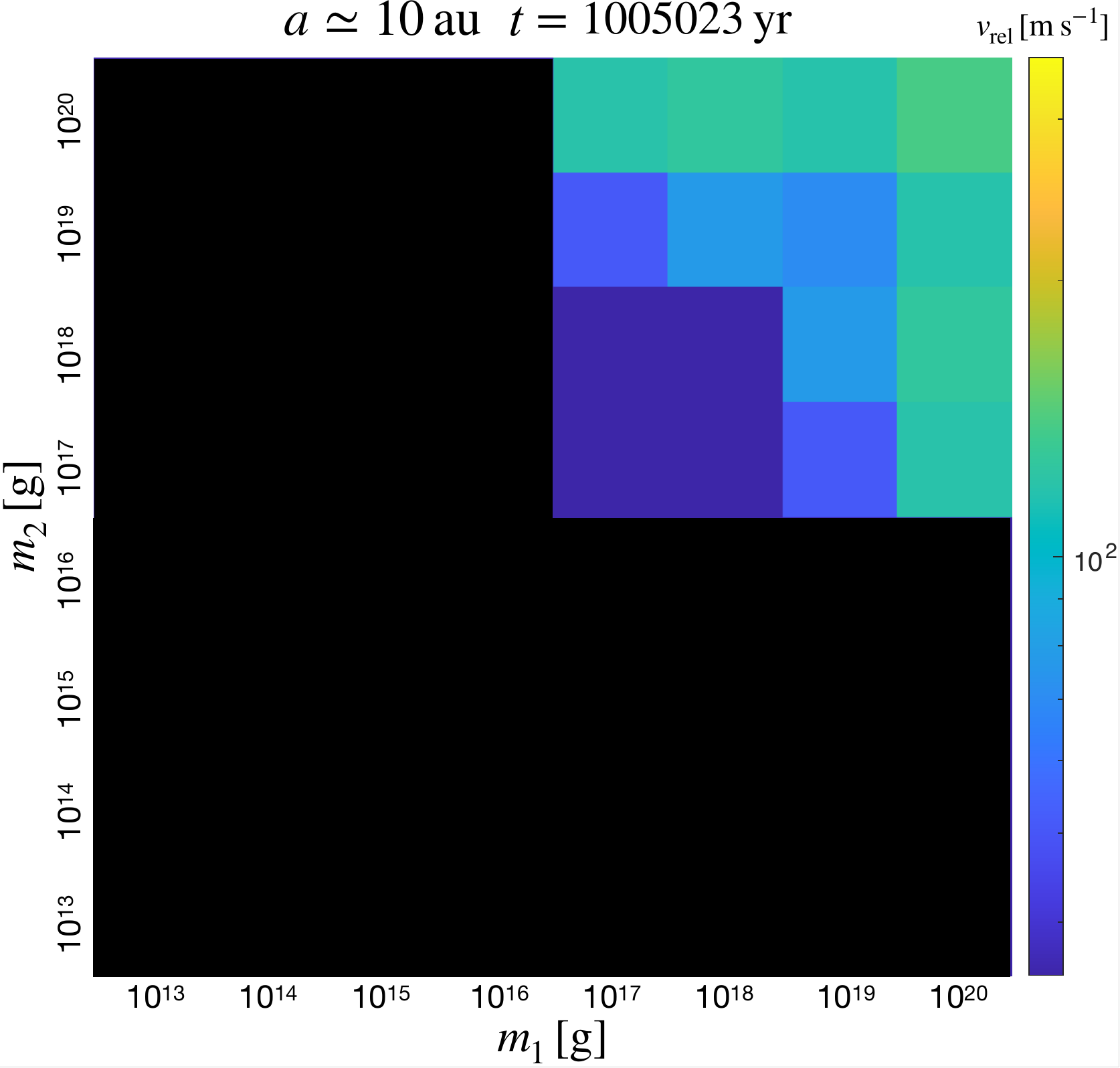}{0.3 \textwidth}{(d)}
	\fig{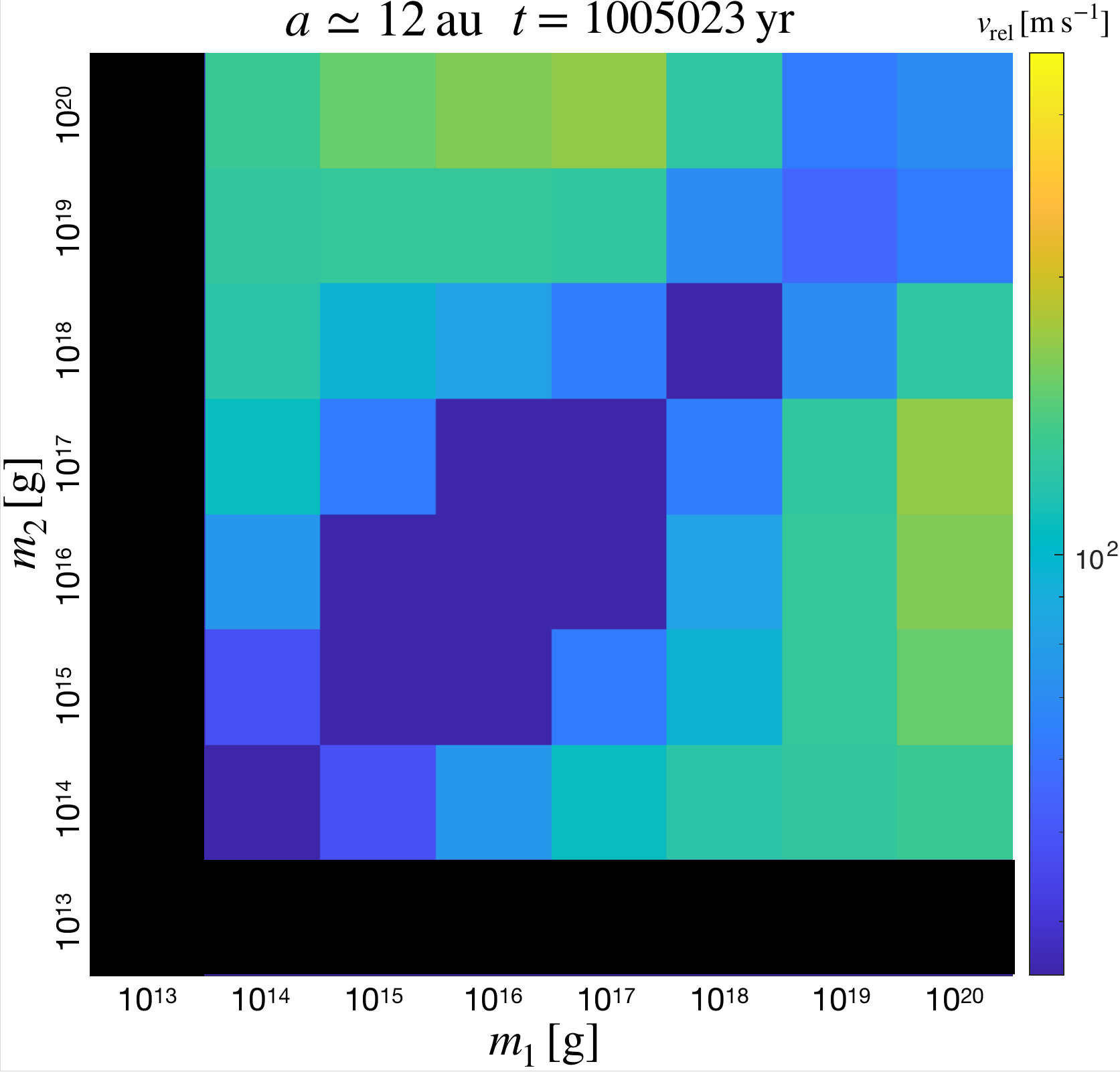}{0.3 \textwidth}{(e)}
	\fig{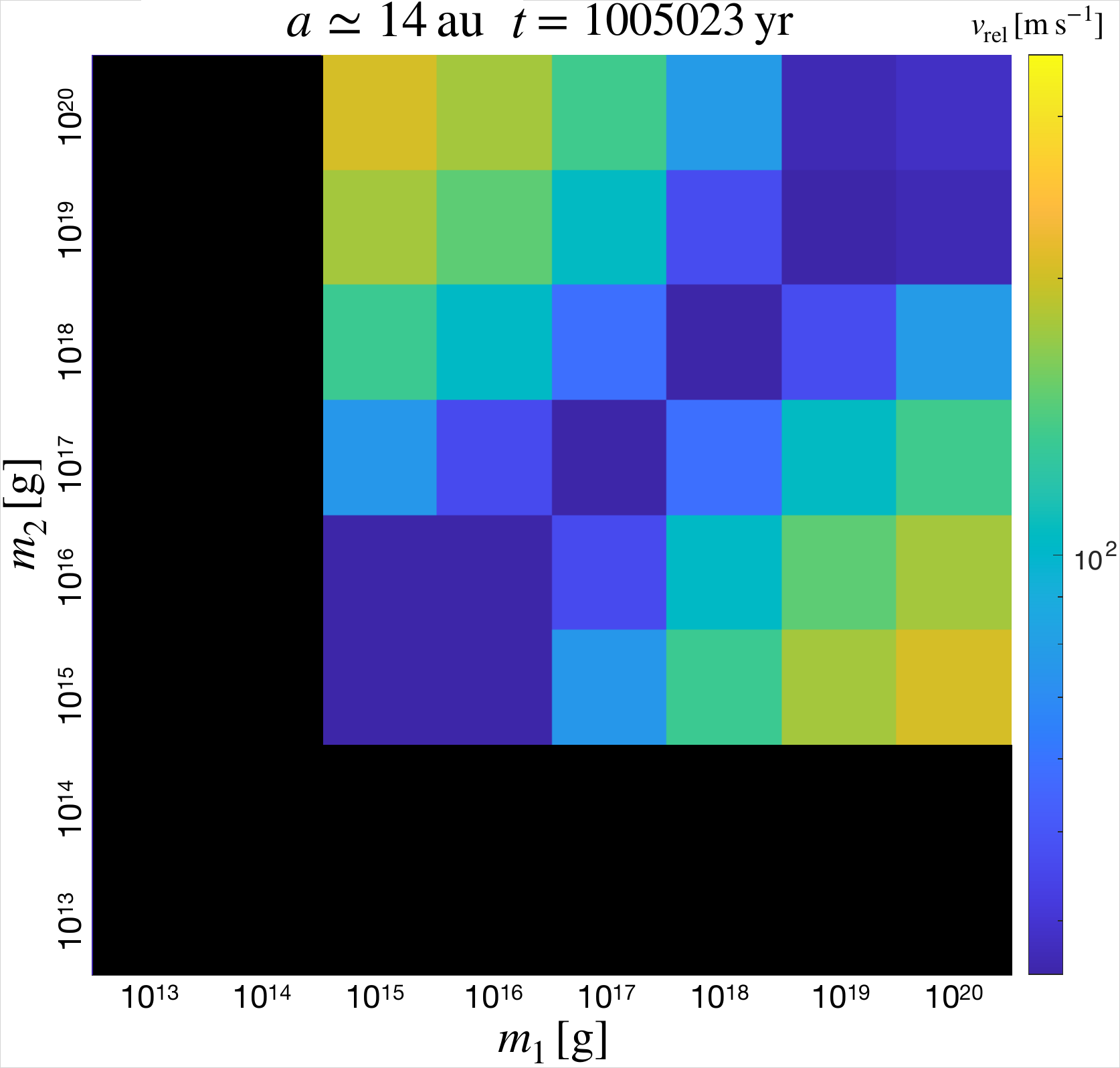}{0.3 \textwidth}{(f)}}

\caption{Color maps of relative velocities at $t \simeq \tau_{\rm{sec}}$ (top row) and $t \simeq 1$ Myr (bottom row). From left to right, the columns show encounter velocities at increasing orbital distance. The black regions are missing data points due to deficiencies of particles at the given range of the semi-major axis.}
\label{fig:v_rel}
\end{figure*}

Fig. \ref{fig:v_rel} shows the color maps of encounter velocities for all particle masses. 
The black regions in the plots mark deficiencies of data points due to orbital decay of particles. 
The snapshots at $a \simeq 12$ au and $a \simeq 14$ au do not suffer from significant loss of data because of the shepherding of particles at the 3:1 and 4:1 resonances.

\subsubsection{Dependence on semi-major axis}

The encounter velocities are high at the MMR, where the eccentricities are excited and the longitudes of pericenter are randomly distributed, as shown in Fig. \ref{fig:v_disp} and Fig. \ref{fig:alignment_oe}. 
Elsewhere, the encounter velocities depend on the degree of alignment of the pericenters.
Take Fig. \ref{fig:v_disp} for example. 
The velocity dispersion (random encounter velocity) among particles with $m = 10^{20 \,}$g can be as high as $\simeq 2500 \, \text{m s}^{-1}$  at the 2:1 MMR (near 8 au). 
Between the 2:1 and 3:1 MMR ($\simeq$ 8-11 au), the velocity dispersion is typically $\sim \, \mathcal{O}(10^2) \,\text{m s}^{-1}$; between 3:1 and 4:1 MMR ($\simeq$ 11-13 au) and outside the 4:1 MMR ($> 13$ au), the velocity dispersion is typically $\sim \, \mathcal{O}(10) \, \text{m s}^{-1}$ (ignoring the higher order MMRs).

From left to right, the columns in Fig. \ref{fig:v_rel} show encounter velocities at increasing orbital distance. 
For identical-mass particles, the encounter velocity (in the diagonals) decreases as $a$ increases. 
This trend generally remains the same elsewhere in the color maps for particles of different masses. 
This is consistent with what is indicated by Fig. \ref{fig:ecc_vec_mid}, discussed in Section \ref{sec:pericenter_a}.
This trend is a result of weaker secular perturbation as well as weaker gas drag at larger orbital distances. 
As a result of orbital decay due to gas drag, some particles (usually those of small masses) are missing near the location where we sample the data, especially at later times in the simulation.
This is also why we do not show the results of encounter velocities at the end of the simulation - the time evolution of the results is not significant and the data set is not complete. 

\subsubsection{Dependence on particle mass}

As discussed in Section \ref{sec:pericenter_m}, smaller particles have more aligned orbits. 
Therefore, smaller particles (of identical mass) generally have lower encounter velocities.
This can be seen from the darker color at the bottom-left end of the diagonals on the color maps in Fig. \ref{fig:v_rel}. 
For a pair of particles (projectile and target), a larger mass ratio leads to a higher encounter velocity, as found in previous studies (e.g., \citet{KW00} and \citet{Kortenkamp_et_al_2001}, \citet{Scholl_et_al_2007}).
The transient preferable feature discussed in Section \ref{sec:pericenter_m}, that particles of larger masses have smaller $\Delta e$ on the $e_x$-$e_y$ plane, seems less significant on the velocity color maps. 
Only plots (b), (c), and (f) show visible decreases of relative velocities between particles of larger masses (darker top-right corner). 
One conclusion we can draw from these results is that, in the disk regions between 8 and 15 au (excluding the MMR locations), the random encounter velocity among particles of identical masses is typically $\sim \mathcal{O}(10) \, \text{m s}^{-1}$, which is significantly lower than that among particles on non-aligned orbits (see MMR locations in Fig. \ref{fig:v_disp}).

\subsection{Acceleration of planet growth}

\begin{figure}
\centering
\label{fig:compare_timescale}
\includegraphics[scale=0.5]{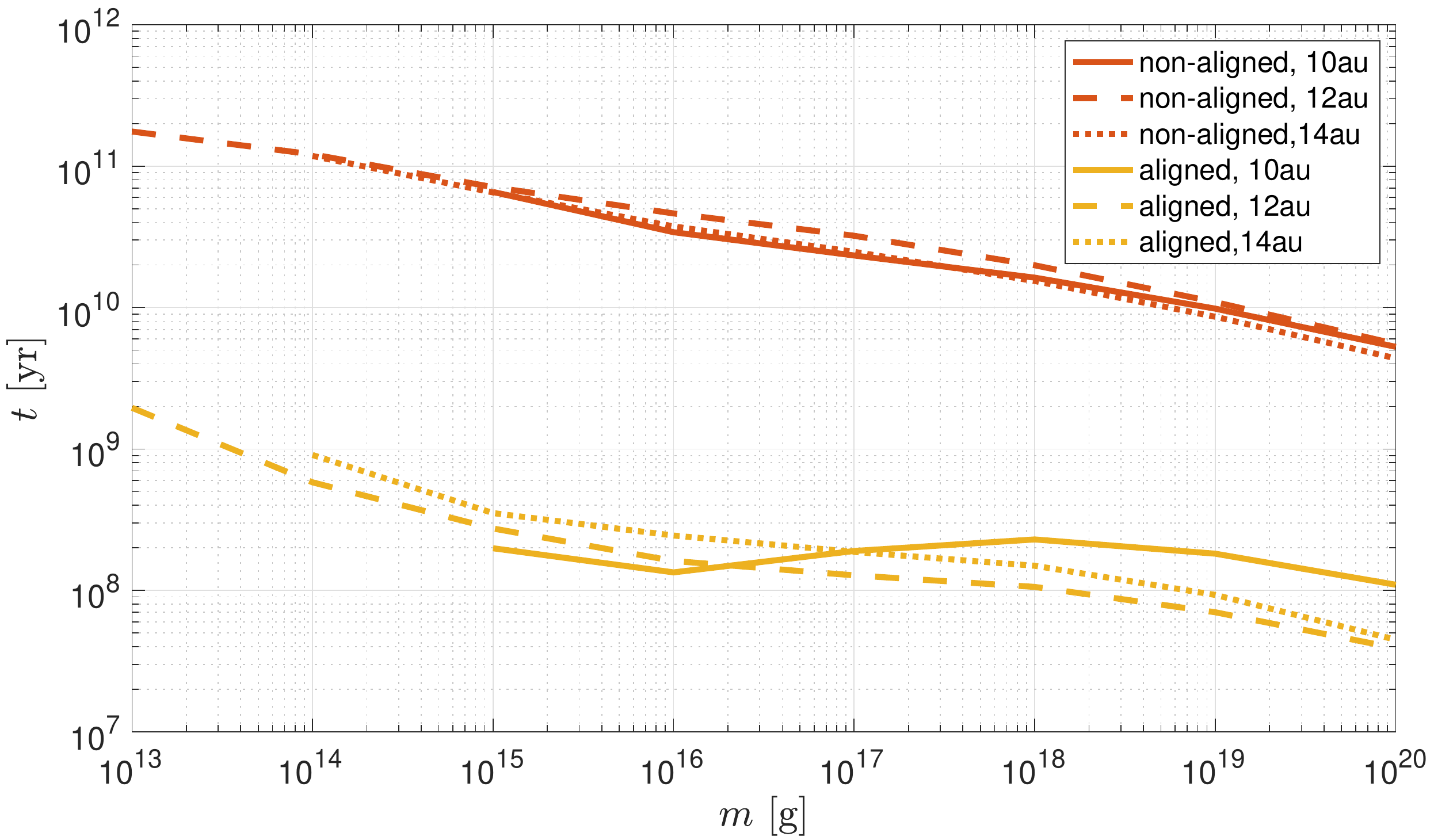}
\caption{Comparison of growth timescales in the aligned case (yellow) and non-aligned case (red) at 10, 12, and 14 au (solid, dashed, and dotted curves). The missing parts of the curves are due to shortages of particles caused by inward drift.}
\label{fig:timescale_compare}
\end{figure}

Although in this work we do not follow the accretionary process of planetesimals, we can quickly examine the effect of pericenter alignment on planet growth by comparing the growth timescales. 
Fig. \ref{fig:compare_timescale} shows rough estimates of the growth timescales in both the aligned and non-aligned cases at three orbital locations. 
We use Eq. (\ref{eq:growth_timescale}) in the Appendix and data from our simulations to calculate the growth timescale for each mass. 
To calculate the term $\langle \tilde{e} ^2 \rangle$ in the equation, we use  $\sigma_e$  in the aligned case and $\langle e \rangle ^{1/2}$ in the non-aligned case. 
It is easy to see that pericenter alignment greatly reduces the growth timescale (for at least one order of magnitude).

\section{Summary and Discussion}  \label{sec:summary}

\subsection{Conclusions}

We investigated planetesimal dynamics under the coupling effect of gas drag and perturbation from a giant planet $M_{\rm{p}} = 1 M_{\rm{Jup}}$ located at 5.2 au in the protoplanetary disk. 
Particularly, we explored the alignment of orbits and its impact on the encounter velocity of planetesimals in the outer region of the disk. 
Our conclusions can be summarized as follows:

\begin{itemize}

\item The outer region of the disk ($a \simeq 9$-15 au), apart from the MMR locations, generally shows patterns of orbital (pericenter) alignment of planetesimals. 

\item The further from the planet, the more well aligned the orbits are, as a result of weaker perturbation. 
Particles of lower mass have better aligned orbits. 

\item Pericenter alignment leads to low random encounter velocity. 
The typical random encounter velocity among identical-mass particles (velocity dispersion) is  $\sim \mathcal{O}(10) \, \text{m s}^{-1}$.

\item The encounter velocity decreases with increasing distance from the perturber and decreasing mass ratio.  

\item When $t$ is sufficiently long (longer than $\tau_{\rm{gas}}(m)$), the pericenters approach their equilibrium locus as mass increases (see Fig. \ref{fig:hk_eq}), resulting in lower relative velocity between larger-mass particles.
\end{itemize} 

Our results suggest that the existence of a giant planet does not necessarily generate high encounter velocity in planetesimals in the disk, which would impede the growth of other planetary bodies. Instead, its secular perturbation together with nebula gas drag leads to low relative velocities in certain regions of the disk (e.g., $\simeq 10$, 12, and 14  au in our model). With the relative velocities held low in the outer region of the disk, the formation of the core of a second giant planet exterior to the perturber planet can be accelerated. 
A more quantitative estimate of the extent to which the core formation is accelerated requires simulations of longer timescale to account for the accretionary processes of planetesimals.

\subsection{Resonance trapping and planetesimal accretion}

Although we try to avoid the MMR locations when displaying our results, the pile-up of particles at those locations, or resonance trapping, is a noteworthy feature. 
When the particles outside the planet drift inward and interact with the nebula gas, they can be halted at resonances by the competing forces of the gravitational kick from the planet and gas drag reaching equilibrium. 
In Fig. \ref{fig:time_evolution}, we can see resonance trapping from vertical strip-like structures in the panels of the first two rows, and bumps in the surface density distribution in the panels of the third row. 
Most significantly pile-ups occur near 8 au (2:1) and near 11 au (3:1). 

Like the random encounter velocity, the solid surface density is also a crucial factor that determines the growth rate of planetary bodies in a certain location. 
One natural question is whether a high solid surface density at the resonances will result in accelerated growth of planetary cores there.
\citet{Beauge_et_al_1994} explored a very similar scenario as that in our paper, but instead of focusing on the alignment of orbits and relative velocity, they followed the whole accretionary process of a planetary core exterior to a Jupiter-like planet. 
With the accumulation of planetesimals at the resonances, they were able to produce a planetary core with orbital elements well in accord with Saturn in a million-year timescale. 
Despite  this encouraging result, it is worth noting that the treatment of gas drag might not be very accurate in their model. 
For planetesimals with a mass of $\sim 10^{25} \,$g , they calculated the gas drag as if their radius was $\sim 10^2$-$10^4\,$m. 
This unnaturally strong gas drag would lead to an underestimate of the eccentricity and, consequently, the encounter velocity of planetesimals trapped in resonances. 

According to our results, although the solid surface density is indeed high at the resonances, the encounter velocity is also high there due to the high eccentricity and non-aligned orbits (see Fig. \ref{fig:time_evolution} and Fig. \ref{fig:alignment_oe}). 
For example, in Fig. \ref{fig:v_disp}, the random encounter velocity of particles with $m = 10^{20} \,$g can be as high as $\simeq 2500\,$m/s at the 2:1 MMR and $\simeq 500 \,$m/s at the 3:1 MMR. 
According to \citet{THEBAULT2006193}, such high encounter velocities would result in disruption or erosion, instead of accretion (see \textit{Large planetesimals case} in \citet{THEBAULT2006193}). 
Therefore, despite the high solid surface density at the resonances, these locations are not quite friendly for planetesimal accretion.

\subsection{Comparison with previous works}

Our work is inspired by previous works to a great extent. 
In particular, the idea of an early-formed gas giant perturbing the planetesimals in the disk origins from \citet{KW00} and \citet{Kortenkamp_et_al_2001}, although they focused on the planetesimals inside the planet. 
The configuration of our model and some of our results also resemble previous studies on circumbinary planetesimal accretion such as \cite{Moriwaki_Nakagawa_2004}, \citet{Scholl_et_al_2007}, and \citet{Marzari_et_al_2008}. 
Here we make a brief comparison between these works and our results, in an effort to find the connections between our model and the circumbinary configuration.

One prominent feature we find, i.e. the size-dependent alignment of the planetesimal orbits, has been identified in almost all the works mentioned above (except for \cite{Moriwaki_Nakagawa_2004} which does not consider gas drag). 
The consequential low encounter velocity among identical-size planetesimals and its effect on planetesimal accretion has been addressed as the Type II runaway growth, where the effect of dynamical friction in the classical Type I runaway growth is mimicked by the size-dependent phasing of orbital elements \citep{Kortenkamp_et_al_2001}. 
This means that this feature is true regardless of whether the planetesimals are inside or outside the orbit of the massive perturber. 

In this study, we fix the mass and the orbit of the planet. 
In other words, we do not explore the dependence on the semi-major axis, the eccentricity, or the mass ratio between the central star and the perturber. 
In \citet{Scholl_et_al_2007}, where the relative velocities among accreting planetesimals in a circumbinary disk are investigated, they fixed the binary separation and probed the dependence of the critical radial distance from the binary barycenter beyond which planetesimal accretion is possible on the binary eccentricity and mass ratio. 
They found that the critical orbital distance is smallest for equal-mass binaries on almost circular orbits, and shifts to larger values for increasing binary eccentricity $e_{\rm{B}}$ and decreasing mass ratio $q$.
They stated that this mass ratio dependence is a natural outcome from the equation of the forced eccentricity given in \citet{Moriwaki_Nakagawa_2004}, where the origin of the coordinate system is located at the barycenter of the binary system:
\begin{equation}
e_{\rm{forced}} = \frac{5}{4}(1-2q)\frac{a_{\rm{B}}}{a}e_{\rm{B}} \Big[\frac{1+(3/4)e_{\rm{B}}^2}{1+(3/2)e_{\rm{B}}^2} \Big].
\end{equation}
Here $q$ is the mass ratio between the primary and the companion ($M_1 = 1-q$, $M_2 = q$); $a_{\rm{B}}$ and $e_{\rm{B}}$ are the binary separation and eccentricity, respectively. 
For the equal-mass binary case ($q = 0.5$), the forced eccentricity becomes 0 and the planetesimal eccentricities are only pumped up by short-period gravitational kick.
Similarly, they found that the relative velocities generally increase as the binary eccentricity increases (see Fig. 2 in \citet{Scholl_et_al_2007}). 

In our model, however, we use a coordinate system with the origin located at the central star. 
Therefore, the equation of the forced eccentricity has a different form as shown in Eq. (\ref{eq:forced_eccentricity}).
In this case, $e_{\rm{forced}}$ increases with increasing planet eccentricity and orbital distance, but it does not depend on the mass of the planet.
Therefore, we cannot infer the consequences of varying the mass of the planet in the same manner as that in \citet{Scholl_et_al_2007}.
All we can possibly speculate is that if the planet becomes more massive, the short-period gravitational kick it exerts on the planetesimals during each encounter becomes stronger so that the dispersion of the eccentricity vectors $\sigma_e$ of an identical-mass particle group increases, resulting in less aligned orbits and increased encounter velocities.

Unlike on the mass of the perturber, the forced eccentricity in our case has a clear dependence on the eccentricity of the perturber as in the case of \citet{Scholl_et_al_2007}.
From Eq. (\ref{eq:forced_eccentricity}) we can see that the forced eccentricity increases monotonically with increasing planet eccentricity. 
With the forced eccentricity increased, the eccentricities of the planetesimals get pumped up by both secular perturbation and short-period kicks, which leads to higher possibility of orbit crossing and thus collisional events. 
However, with gas drag in presence, the size-dependent phasing of orbits helps reducing the relative velocities of planetesimals even with high eccentricities. 
Therefore, instead of analytically predicting that the relative velocities would increase with increasing planet eccentricity, we can only infer from the numerical results in \citet{Scholl_et_al_2007} that the increase of the perturber's eccentricity would result in higher relative velocities of planetesimals at a given radial distance in the disk. 

The dependence on the semi-major axis of the planet $a_{\rm{p}}$ can be inferred easily.
Although we fix $a_{\rm{p}}$ at 5.2 in our model, it is easy to infer from our results at different $a$ that an increase in $a_{\rm{p}}$ would generally result in higher relative velocities of planetesimals in a given range of the disk (exterior to the planet), and vice versa.

It is difficult to give precise quantitative estimates of how the situation changes when we vary the parameters of the planet (in particular, $m_{\rm{p}}$ and $e_{\rm{p}}$) without actually conducting numerical simulations. 
We can only speculate that larger distance from the planet, smaller eccentricity, and smaller mass of the planet would probably be more favorable for remaining low encounter velocities among planetesimals, which is preferable for accretion.

\subsection{Limitations and future perspectives}

One potential problem of this work is that we neglect the mutual gravitational interactions of the planetesimals. 
According to our estimate, the timescale for viscous stirring for most particle masses is much longer than that for gas drag and secular perturbation. 
However, it shortens drastically as the particle mass increases. 
It is possible that the mutual perturbation between planetesimals would have a significant impact on our results, especially for particles with large masses. 
For example, while secular perturbation and gas drag work to maintain a size-dependent orbital alignment, the self-gravity of planetesimals might tend to disrupt such alignment by increasing the random velocity.
It is unclear how and to what extent the self-gravity would change the relative velocities of planetesimals in the parts of the disk we consider.

Another limitation is that we do not consider the possible migration of the perturber throughout the process.
The effect of secular perturbation sensitively depends on the distance between the perturber and the particle. 
As noted in \citet{KW00}, the Jupiter-Saturn perturbations would be considerably smaller if Jupiter and Saturn had larger semi-major axes at the time of planet formation, and drifted inward to their present positions. 
Radial migration of giant planets is commonly believed to be caused by gravitational interactions between the massive planet and the gas disk \citep{goldreich1980disk, ward1989orbital, KW00}. 
Had the perturber planet in our case drifted inward while we simulated the orbital evolution of the planetesimals, the time evolution and spatial distribution of orbital elements,and the relative velocities might be considerably different from our current results. 

A full {\em N}-body simulation including the self-gravity of planetesimals is currently underway.
We intend to modify the code so that we can run the integrator on GPUs to accelerate the calculation. 
The migration of the perturber is a tricky problem because it is difficult to track the migration through planet-disk interactions with hydrodynamic simulations at the same time as {\em N}-body simulations. 
However, it is possible to use analytical prescriptions for the radial migration of a planet, and add the heliocentric distance of the perturber as a variable in our simulation. 

Finally, it is noteworthy that the framework of this study is also applicable to exploring planetesimal dynamics in binary star systems. 
For example, the formation of planets in a circumbinary disk (P-type) is an interesting but poorly understood topic. 
In future work, we will apply our results to planetesimal dynamics in a circumbinary disk.

\acknowledgements

The Numerical computations were carried out on the general-purpose PC cluster at the Center for Computational Astrophysics, National Astronomical Observatory of Japan.
E.K. is supported by JSPS KAKENHI Grant Number 18H05438.

\bibliography{draft}
\bibliographystyle{aasjournal}

\end{document}